\documentclass[10pt,letterpaper]{article}
\usepackage[top=0.85in,left=2.75in,footskip=0.75in]{geometry}

\usepackage{amsmath,amssymb}

\usepackage{changepage}

\usepackage{textcomp,marvosym}

\usepackage{cite}

\usepackage{nameref,hyperref}

\usepackage[right]{lineno}

\usepackage[nopatch=eqnum]{microtype}
\DisableLigatures[f]{encoding = *, family = * }

\usepackage[table]{xcolor}

\usepackage{array}

\newcolumntype{+}{!{\vrule width 2pt}}

\newlength\savedwidth

\raggedright
\usepackage[aboveskip=1pt,labelfont=bf,labelsep=period,justification=raggedright,singlelinecheck=off]{caption}

\makeatletter
\renewcommand{\@biblabel}[1]{\quad#1.}
\makeatother

\usepackage{lastpage,fancyhdr,graphicx}
\usepackage{epstopdf}
\fancyheadoffset[L]{2.25in}
\fancyfootoffset[L]{2.25in}
\usepackage{graphicx}%
\usepackage{multirow}%
\usepackage{amsmath,amssymb,amsfonts}%
\usepackage{amsthm}%
\usepackage{mathrsfs}%
\usepackage[title]{appendix}%
\usepackage{xcolor}%
\usepackage{textcomp}%
\usepackage{manyfoot}%
\usepackage{booktabs}%
\usepackage{algorithm}%
\usepackage{algorithmicx}%
\usepackage{algpseudocode}%
\usepackage{listings}%
\usepackage{adjustbox}

\usepackage[utf8]{inputenc}

\usepackage{xcolor} 
\usepackage{graphicx} 
\usepackage{subcaption} 
\usepackage{amsmath} 
\usepackage{verbatim} 
\usepackage{hyperref} 
\usepackage{caption} 
\usepackage{subcaption} 
\usepackage{comment} 

\usepackage{tikz}  
\usetikzlibrary{matrix}  
\usetikzlibrary{positioning}
\usetikzlibrary{arrows.meta}
\usetikzlibrary{decorations.pathreplacing}
\usetikzlibrary{calc}

\DeclareMathOperator{\maxthetaX}{\hat{\theta}_{\max}(\mathbf{X})}
\DeclareMathOperator{\maxtheta}{\hat{\theta}_{\max}}

\begin{document}
\vspace*{0.2in}

\begin{flushleft}
{\Large
\textbf{Inventing a coin-flip classifier: Accounting for multiplicity in machine learning benchmark performance} 
}
\newline
\\
Kajsa Møllersen\textsuperscript{1,2*},
Einar Holsbø\textsuperscript{2}
\\
\bigskip
\textbf{1} Sámi University of Applied Sciences, Guovdageaidnu, Norway
\\
\textbf{2} Department of Computer Science, UiT - The Arctic University of Norway, Tromsø, Norway
\\
\bigskip

*kajsam@samas.no

\end{flushleft}

\section*{Abstract}
State-of-the-art (SOTA) performance refers to the highest performance achieved by some model on a test sample, under controlled conditions such as public data (reproducibility) or public challenges (independent sample). Thousands of classifiers are applied, and the highest performance becomes the new reference point for a particular problem. In effect, this set-up is an estimate of the expected best performance in a classifier ensemble applied to a random sample; a sample maximum estimate. In this paper, we argue that SOTA should instead be estimated by the expected performance of the best classifier, which can be done without knowing which classifier it is. 

Time and time again, we have seen a significant drop in performance when best performing models are applied to new data samples, even when the original result was achieved on an independent sample. 
The reason for this, is that picking the highest performance from an ensemble of classifiers is in effect a sample maximum estimate, whereas applying that one best performing classifier to a new data sample gives an estimate of a single classifier's performance. 
The former is a biased estimator of the latter, and the performance drop is unavoidable. 

In other words, if you pick the highest observed performance created by an ensemble of classifiers applied to a data sample, this number will drop, often significantly, when you apply that single classifier to a new data sample.
The bias reflects a multiplicity effect. 

Our contribution is the formal distinction between the two, and an investigation into the practical consequences of using the former to estimate the latter.
This is done by presenting sample maximum estimator distributions for non-identical and dependent classifiers.
We illustrate the impact on real world examples from public challenges.


\section*{Introduction} 
\label{sec:Introduction}

Machine learning (ML) models are commonly evaluated by their observed performance on data sets from public or private repositories, where models can be compared under identical conditions and across time. 
For the last couple of decades, the success of more complex models has increased the use of observed performance on benchmark data sets as evidence.
We direct attention to the use of observed performance as state-of-the-art (SOTA) reference point for classifiers, but the overall mechanisms apply to other models. 
Other performance measures, such as computation time, will not be addressed. 

Within the fields of research that develop, refine, test and use classifiers, highest reported observed performance is used as a reference point, see, for example paperswithcode.com, and is commonly referred to as SOTA.
But the highest observed performance is only an estimate of SOTA, and a biased one as such, due to the inherent multiplicity. 
Multiplicity is, simply put, how the probability of an outcome changes when an experiment is performed multiple times.
Many aspects of public benchmark datasets and public challenges have been thoroughly discussed in the literature, and we present some of the topics connected to multiplicity here.

\subsection*{Related work} \label{sec:RelatedWork}

Multiple comparisons is a well-studied subject within statistics, see, for example,  
\cite{Tukey1991} for an interesting discussion on some topics, among them 
confidence intervals (CIs). 
For a direct demonstration of how multiple testing can lead to false discoveries, \cite{Bennett2009a} gave an example where brain activity is detected in (dead) salmon ``looking'' at photos. 
In the ML context, with multiple data sets and classifiers, \cite{Demsar2006} 
suggested adjusting for multiple testing, according to well-known statistical theory. 
This paper focuses on the estimate itself and its uncertainty, and not comparison of methods. 

The use and re-use of benchmark datasets
have been addressed in several publications. 
\cite{Longjohn2025Benchmark} drew attention to the role of data repositories with leaderboards and their lack of metric uncertainty. 
\cite{Salzberg1997} pointed to the problem that arises because these datasets are static, and therefore run the risk of being exhausted by overfitting. 
Exhausted public data sets still serve important functions, for example, a new idea can show to perform satisfactory, even when it is not superior.
\cite{Thompson2020} 
distinguished between simultaneous and sequential multiple correction procedures. 
An interesting topic they bring up is the conflict between sharing incentive, open access and stable false positive rate. 
There are several demonstrations of the vulnerability of re-use of benchmark datasets. 
\cite{Teney2020} performed a successful attack on the VQA-CP benchmark dataset \cite{Agrawal2018}; they were able to attain high observed performance with a non-generalizable classifier.
They point towards the risk of ML methods capturing idiosyncrasies of a dataset.
\cite{Torralba2011} introduced the game Name That Dataset! motivating their concern of methods being tailored to a dataset rather than a problem. 
The lack of statistically significant differences among top performing methods has also been addressed \cite{Everingham2015, Fernandez2014}. 

The same mechanisms of overfitting are present in public challenges through multiple submissions on the validation sample. 
\cite{Blum2015} 
introduced an algorithm that releases the public leaderboard score only if there has been significant improvement compared to the previous submission. 
This will prevent overfitting by hindering adjustments to random fluctuations in the score and not true improvements of the method.
\cite{Dwork2017} used the principles of differential privacy to prevent overfitting to the validation set. 

Several publications show that the overfitting was not as bad as expected, and provide possible explanations. 
\cite{Recht2019} constructed new, independent test sets of CIFAR-10 and ImageNet. 
They found a dramatic drop in accuracy, around 10\%. 
However, the order of the best performing methods is preserved,
which can be explained by the fact that
multiple classes \cite{Feldman2019} and model similarity \cite{Mania2019} slow down overfitting.
\cite{Roelofs2019} compared the results from the private leaderboard (one submission per team) and the public leaderboard (several submissions per team) in Kaggle challenges.
Several of their analyses indicate overfitting.

Overfitting is 
avoided in most challenges by withholding the test sample so that a method can only be evaluated once and not adapted to the result.
Hold-out datasets also prevents wrong use of cross-validation \cite{Fernandez2014, Dietterich1998}.
Another contribution from public challenges is the requirement of reproducibility to collect the prize money, which is a well-known concern in science \cite{Gundersen2018}. 
Public challenges typically rank the participants by a (single) performance measure, and can fall victims of Goodhart's law, paraphrased as ``When a measure becomes a target, it ceases to be a good measure.'' \cite{Teney2020}.
\cite{Mollersen2017} demonstrated how five different performance measures completely rearranges the leaderboard for the top participants in the ISBI 2016 challenge Skin Lesion Analysis towards Melanoma Detection \cite{Gutman2016}.
\cite{Ma2021} offered a framework and platform for evaluation of NLP methods, where other aspects such as memory use and robustness are evaluated.

The hold-out test set 
does not offer an immediate solution the
robustness of a method and the variability in observed performance. 
Choice of hyperparameters are often adjusted to validation set results, but can be hindered \cite{Blum2015, Dwork2017} 
if implemented by the challenge hosts. 
\cite{Bousquet2002} investigated how sampling randomness influences accuracy, both for regression algorithms and classification algorithms, relying on the work of \cite{Talagrand1996} who stated that ``A random variable that depends (in a ``smooth'' way) on the influence of many independent variables (but not too much of any of them) is essentially constant.''
The mechanisms that create the overestimation of SOTA is closely related to regression toward the mean \cite{Galton1886Regression} and winner's curse \cite{Capen1971Competitive}. 
Both are related to extreme value statistics, in the sense that the effect of these phenomena are major when extreme values appear, as both are concerning sample maxima. 
However, regression toward the mean and winner's curse are not eradicated in the absence of extreme values. 
We take a look at SOTA estimation in this context, but from a slightly different angle and set-up. 

There is a rich literature on various aspects regarding the use of public challenges as standards for scientific publications, see, for example, \cite{Varoquaux2022}, for a much broader overview than the one given here.

\subsection*{Outline and contributions} \label{sec:Outline}

There is an important distinction between best observed performance in an ensemble of classifiers, and expected performance of best classifier, and this paper brings attention to that.

In Section~\nameref{sec:Definitions}, we define SOTA classifier performance (Definition~\ref{def:SOTA}), so that it can be clearly distinguished from the sample maximum estimation (Definition~\ref{def:maxthetaX}). 
With this in place, we present its distribution function and bias (Section~\nameref{sec:Multiple}), and suggest a heuristic method for unbiased SOTA estimation (Section~\nameref{sec:estimate_SOTA}). 
We demonstrate its impact on simulated and real world data for validation in Section~\nameref{sec:Simulations}, and discuss the utility of these contributions in Section~\nameref{sec:Discussion}. 
All code is available at https://github.com/3inar/ninety-nine.

\section*{Definitions} \label{sec:Definitions}


The term \textit{performance} refers to whatever measure is used to rank or quantify a model's ability to fulfill its task; typically the accuracy of a classifier. 
We refer to the expected performance of the best model as SOTA, and distinguish it from the best observed performance, which is a biased estimate of the former. 
Note that the expected performance, corresponding to the ``true, underlying" performance of a model, is usually unknown, and commonly inaccessible for complex models. 
We use the term \textit{performance} referring to the inherit performance of a model, and \textit{observed performance} for the specific observation on a data sample. 

Classifiers are commonly not trained before the challenge is announced and the training sample is released.
\cite{Bouthillier2021}
showed empirically that hyperparameter choice and the random nature of the data are two large contributors to variance, whereas data augmentation, dropouts, and weights initialisations are minor contributors. 
Team participation is yet another random factor, similar to the company participation in a bid, modelled as probabilities in \cite{Capen1971Competitive}.
The random model also coincides with empirical Bayes approach \cite{Efron1973Stein}. 

We therefore model the classifier ensemble as a random sample, 
and define SOTA performance as follows: \\

\noindent \textbf{Definition 1 (SOTA classifier performance)}
Let $\theta_j$ be the random variable that denotes the performance of classifier $j$.
In an ensemble of $m$ classifiers with performances $\Theta = \{\theta_1, \theta_2, \ldots, \theta_m\}$, 
the SOTA classifier performance is the expected best performance 
\begin{align}
    \theta_{SOTA} = \text{E}\max \Theta.
    \label{def:SOTA}
\end{align}

\noindent 
This definition answers the question \textit{What can we expect the best classifier's performance to be?} when applied to a new data sample.
Note that picking the highest observed performance derived from an ensemble of classifiers applied to a test sample does not answer this question. 

The observed performances, $\hat{\theta}_j$'s, based on the application of models on a specific benchmark or test sample, are estimates. 
The test samples act as random samples, all the while the point of reporting these observed performances is the assumption that they can be generalized to similar data sets. 
Hence, the $\hat{\theta}_j$'s are a random sample, and the best observed performance is the sample maximum estimate: \\

\noindent \textbf{Definition 2 (Sample maximum estimator)}
Let $\theta_j$ be the random variable that denotes the performance of classifier $j$ and let $\hat\theta_j(X)$ be the corresponding unbiased estimator. 
For an ensemble of $m$ classifiers with performances $\Theta = \{\theta_1, \theta_2, \ldots, \theta_m\}$, 
the sample maximum performance estimator is 
\begin{align}
    \maxthetaX = \text{E}\max (\hat{\Theta}(\mathbf{X})), 
    \label{def:maxthetaX}
\end{align}
where $\mathbf{X} = \{X_1, X_2, \ldots, X_m\}$ quantifies successes for classifiers $j = 1, \ldots, m$.
The sample maximum answers the question \textit{What is the best expected performance of an ensemble of classifiers?} 
The distinction from Definition~\ref{def:SOTA} can easily be overlooked, but when $\maxthetaX$ is used as an estimator for $\theta_{SOTA}$, it is biased.

\section*{Multiple classifiers and biased SOTA estimation} 
\label{sec:Multiple}

\subsection*{Notation} 



\begin{align*}
    n&: \text{number of trials/size of test sample} \\
    m&: \text{number of experiments/size of classifier ensemble} \\
    Y&: \text{indicator variable for success} \\
    X&: \text{number of successes} \\ 
    \theta = P(X = x) &: \text{probability of $x$ successes}\\
    P_x = P(X\leq x)&: \text{probability of at most $x$ successes} \\
    \mathbf{X} = \{X_1, X_2, \ldots, X_m\}&:\text{number of successes for classifiers }j = 1, \ldots, m 
    \\ 
    \Theta =  \{\theta_1, \theta_2, \ldots, \theta_m\} & : \text{probability of success for classifiers }j = 1, \ldots, m  \\
    & \,\, \,\, \theta \sim P(\theta) \\
    \Theta' =  \{\theta'_1, \theta'_2, \ldots, \theta'_m\} & : \text{empirical Bayes estimate for } P(\theta)  \\
    \hat{\Theta}(\mathbf{X}) = \{ \hat{\theta}_1(X_1), \hat{\theta}_2(X_2), \ldots, \hat{\theta}_m(X_3)\} & : \text{estimator for } \Theta\\
     \theta_{SOTA} = E \max(\Theta) &: \text{expected highest probability of success}\\
    \maxthetaX = \max (\hat{\Theta}(\mathbf{X})) &: \text{sample maximum estimator} \\ 
\end{align*}

We give a motivating example to illustrate the effect of multiplicity.
For any classifier that predicts the outcome of the flip of a fair coin, the probability of success is $0.5$, and hence the SOTA performance is $\theta_{SOTA} = 0.5$. 
In a public coin-flip-classifier challenge, what would the observed highest performance be? 

Consider an experiment where a classifier predicts the outcome of a fair coin being flipped $n=20$ times.
Let the random variable $X$ denote the number of successes, 
and let ${\hat{\theta}(X)=X/n}$. 
We are interested in the probability of observing a large $\hat{\theta}$, say, $\geq 0.90$, and wrongfully getting the impression that a coin-flip classifier has been invented. 
For a single classifier, the probability is 
$0.00020$, reasonably low for making the false claim of  coin-flip classifier invention. 
For the multiplicity setting with $m=1,000$ independent classifiers, 
the probability of at least one classifier performing at $0.90$ or above is $0.18$, a substantial risk of grossly overestimating $\theta_{SOTA}$.

The same mechanism as described here is present when highest observed performance is reported from public challenge leaderboards or public data sets.
Even without the usual suspects of data re-use, overfitting, wrong use of cross-validation, multiple testing or data leakage, multiplicity, when not accounted for, give the impression of the invention of a coin-flip classifier. 

Let $\Theta = \{\theta_1, \theta_2, \ldots, \theta_m\}$ be the probability of success for $m$ classifiers, and let $\mathbf{X} = \{X_1, X_2, \ldots, X_m\}$ be the set of random variables that denotes the number of successes (correct predictions) for each classifier on a sample of size $n$.
Let the statistic $\maxthetaX = \max(\{\hat{\theta}_1(X_1), \hat{\theta}_2(X_2), \ldots, \hat{\theta}_m(X_m)\})$ be the naive sample maximum estimator for $\theta_{SOTA}$.
We are interested in the probability distribution of $\maxthetaX$.

Let $\theta_j \sim P(\theta)$, where the $m$ classifiers from where $\mathbf{X}$ originates are themselves a random sample. 
We can model the conditional distribution as a hierarchical model, and have the following probability mass function (pmf): 
\begin{align}
    \maxthetaX|\Theta &\sim P(\maxthetaX|\Theta, m, n, \rho)  \label{eq:hierarch_X}\tag{1a} \\
    \theta_j & \sim P(\theta) \label{eq:hierarch_theta}. \tag{1b}
\end{align}
where $m, n, \rho$ are fixed, and $\rho$ is the correlation coefficient, see Section~\nameref{sec:id_dep}.
The sample maximum is equivalent to at least one classifier having more than $x$ correct predictions, and can be written as $1-P(\maxthetaX \leq x/n)$, which in turn corresponds to $1 - F(x|\Theta)$. 
In the case where $\Theta$ is fixed the hierarchical model reduces to Eq.~\ref{eq:hierarch_X}, and we denote it simply by $F(x)$.
We refer to classifiers with identical probability of success as identical classifiers. 

For non-identical classifiers, we use $\theta \sim \mathcal{U}(a,b)$, the continuous uniform distribution, for which the sample maximum statistic, $T$, has probability density function (pdf) 
\begin{align*}
    f(t) = \frac{m(t-a)^{m-1}}{(b-a)^m},
\end{align*}
and expected value 
\begin{align*}
    \theta_{SOTA} = E\max (\Theta) = E T = \frac{mb+a}{m+1}.
\end{align*}

Note that the sample maximum estimator is a special case of the order statistic.

\setcounter{equation}{1} 

\subsection*{Identical, independent classifiers} 
\label{sec:id_indep}

We have cumulative distribution function (cdf) 
\begin{align} \label{eq:cdf}
    F(x) 
    &= P_x^{m} = P(X \leq x)^m = \big(\sum_{\ell = 0}^x P(X = \ell) \big)^m,
\end{align}
and the corresponding pmf
\begin{align} \label{eq:pmf}
    f(x) 
    &= P_x^m-P_{x-1}^m = 
    \big(\sum_{\ell = 0}^x P(X = \ell) \big)^m - \big(\sum_{\ell = 0}^{x-1} P(X = \ell) \big)^m \\
    &= \sum_{k = 0}^{m-1}\binom{m}{k}\big( \sum_{\ell = 0}^{x-1}P(X=\ell)\big)^k P(X=\ell)^{m-k}, \nonumber
\end{align}
where $f(x=0) = F(x=0)$. See \cite[p.~228]{Casella2002Statistical} for details. 

\begin{figure}[htb!]
    \centering
    \textbf{Cdf and pdf for identical, independent classifiers}\par\medskip  
    \begin{subfigure}[t]{0.45\textwidth}
      \includegraphics[width=\linewidth]{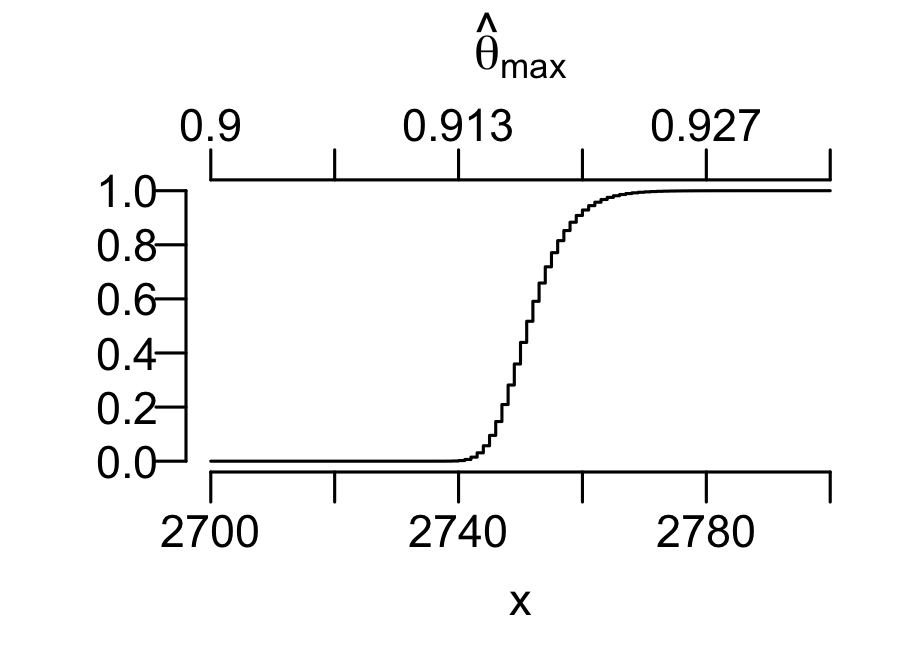}
    \caption{Cdf $F(x)$, see Eq.~\ref{eq:cdf}}
    \label{fig:cumul_success}
    \end{subfigure}
    \hfill
    \begin{subfigure}[t]{0.45\textwidth}
    \centering
   \includegraphics[width=\linewidth]{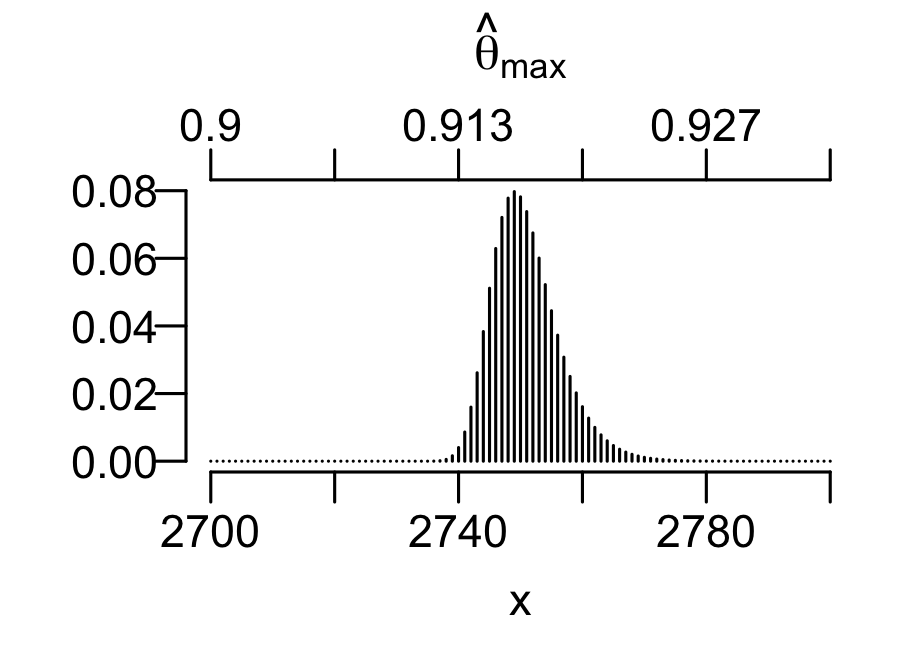}
  \caption{Pmf $f(x)$, see Eq.~\ref{eq:pmf}}
  \label{fig:pmf_success}
  \end{subfigure}
  \caption{The cdf and pmf of the sample maximum estimator for $n = 3,000$, $\theta=0.90$, and $m=1,000$. The horizontal axes denotes the number of successes, $x$, at the bottom and the corresponding sample maximum estimates, $\maxtheta$, at the top. 
  }
  \label{fig:cdf_pmf_success}
\end{figure}

Fig.~\ref{fig:cdf_pmf_success} shows an example of cdf and pmf, and Fig.~\ref{fig:multi_ci} in Section~\nameref{res:id_indep} demonstrates the huge impact multiplicity has on SOTA estimation.

\subsection*{Non-identical, independent classifiers} 
\label{sec:nonid_indep}

To generalize Eq.~\ref{eq:cdf} to the non-identical case,
we seek the sum of $P_{xj} = P(X \leq x|\theta_j)$ to find the proper expression for $F(x|\Theta) = P(C_x = m|\Theta)$.

The Poisson Binomial distribution describes the sum of $m$ non-identical independent Bernoulli trials with success probabilities $p_1, p_2, \ldots, p_m$ \cite{Wang1993}. 
The pmf for $K$ number of successes is
\begin{align*} 
    P(K=k)=\sum \limits _{{A\in G_{k}}}\prod \limits _{{j\in A}}p_{j}\prod \limits _{{i\in A^{c}}}(1-p_{i}),
\end{align*}
where $G_k$ is all subsets of size $k$ that can be selected from \{1, 2, \ldots, m\}.
$G_m$ contains only one subset $A = \{1, 2, \ldots, m\}$, and its complement $A^C$ is empty.

\begin{figure}[t!]
    \centering
    \textbf{Cdf and pmf for non-identical, independent classifiers}\par\medskip 
    \begin{subfigure}[t]{0.45\textwidth}
    \includegraphics[width=\linewidth]{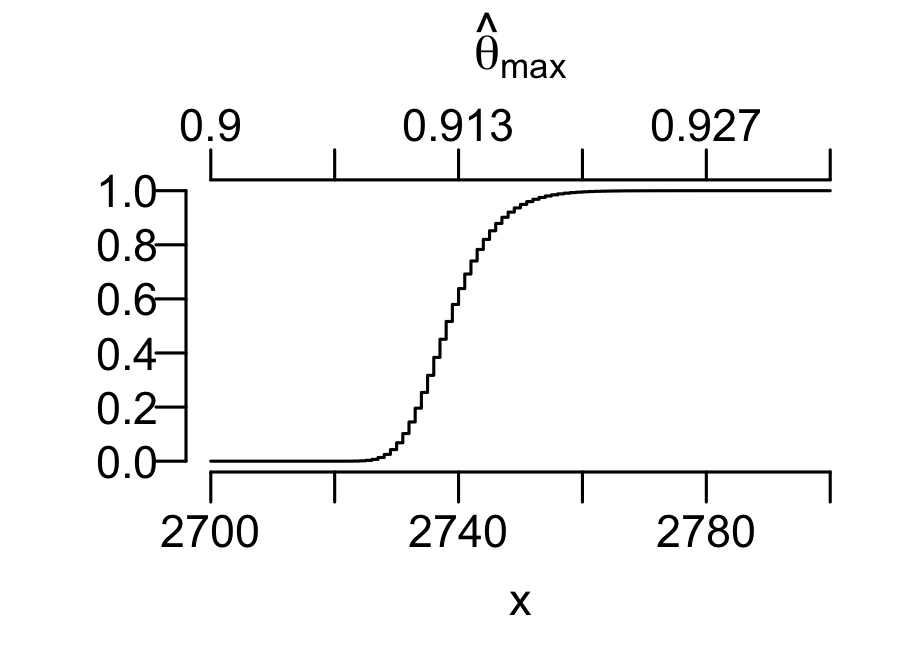}
    \caption{Cdf $F(x)$, see Eq.~\ref{eq:noniid_cdf}}
    \label{fig:noniid_cdf_success}
    \end{subfigure}
    \hfill
    \begin{subfigure}[t]{0.45\textwidth}
    \centering
  \includegraphics[width=\linewidth]{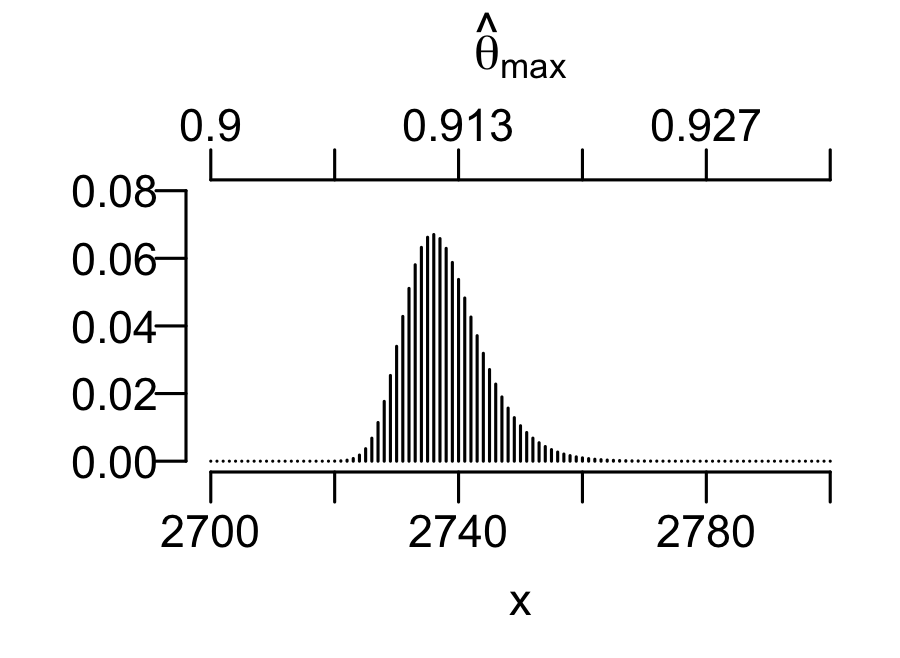}
  \caption{Pmf $f(x)$, see Eq.~\ref{eq:noniid_pmf}}
  \label{fig:noniid_pmf_success}
  \end{subfigure}
  \caption{The cdf and pmf of the sample maximum estimator for $n = 3,000$, $\theta=0.90$, and $m=1,000$ for non-identical $\theta$'s, with $\theta_j \sim \mathcal{U}(a,b)$ where $d = b-a = 0.025$. 
  The horizontal axes denotes the number of successes, $x$, at the bottom and the corresponding sample maximum estimates, $\maxtheta$, at the top. 
  See Fig.~\ref{fig:cdf_pmf_success} for the corresponding distributions for identical classifiers.}
  \label{fig:cdf_pmf_noniid_success}
\end{figure}

We get the following cdf for non-identical classifiers:
\begin{align} \label{eq:noniid_cdf}
    F(x|\Theta) &= \sum \limits _{{A\in G_{m}}}\prod \limits _{{j\in A}}P_{xj}\prod \limits _{{i\in A^{c}}}(1-P_{xi}) 
    = \prod \limits _{j = 1}^m P_{xj}.
\end{align}
The corresponding pmf is
\begin{align} \label{eq:noniid_pmf}
    f(x|\Theta) 
    &= \prod \limits _{j = 1}^m P_{xj} - \prod \limits _{j = 1}^m P_{x\text{-}1j} \\
    &= \prod \limits _{j = 1}^m \sum \limits _{k=0}^{x} \binom{n}{k}\theta_j^k (1-\theta_j)^{n\text{-}k} - \prod \limits _{j = 1}^m \sum \limits _{k=0}^{x\text{-1}} \binom{n}{k}\theta_j^k (1-\theta_j)^{n\text{-}k}. \nonumber
\end{align}
An example of cdf and pmf are displayed in Fig.~\ref{fig:cdf_pmf_noniid_success}. 
Analytically, it is more complex than the case of identical classifiers, but it is easy to simulate, see Section~\nameref{res:nonid_indep}.
\subsection*{Identical, dependent classifiers} 
\label{sec:id_dep}


Even though the different teams participating in a challenge work independently, their resulting classifiers are not necessarily statistically independent. 
In their training phase, different classifiers will pick up on many of the same features and dependencies, regardless of which classifier they train. 
In addition, many teams will use different versions of related models pre-trained on the same data set.
This scenario can be modeled by conditional independency. 

Let $Y_{0}$ be a Bernoulli variable with $P(Y_{0} = 1) = \theta_0$.
Let $Y_{j}$ be the Bernoulli variable with value $1$ if classifier $j$ predicts a data point correctly with $P(Y_{j} = 1) = \theta_j$, and let $j = 1, \ldots, m$ be the $m$ classifiers contributing to $\maxtheta$. 
\cite{Boland1989} described dependency in the majority systems context, where $\rho_{0j}=\text{corr}(Y_0, Y_j), \, j = 1, \ldots m$, and the $Y_j$s are independent given $Y_0$.
$Y_0$ does not contribute directly to $\maxtheta$, but can be seen as, for example, the influence that a common training sample has on the classifiers. 

The binomial distribution for dependent Bernoulli trials is not an easy problem, see, for example, \cite{Ladd1975} and \cite{Hisakado2006}. 
We use a hierarchical model for the pmf as follows: 
\begin{align*}
    \maxthetaX|\Theta, Y_0 &\sim P(\maxthetaX|\Theta,Y_0) \\
    \theta  \sim P(\theta), &\,\,\,Y_0 \sim P(Y_0|\theta_0). 
\end{align*}
The Bernoulli trials can be simulated, see \cite{Boland1989} for conditional distributions.

The dependency model is further discussed in Section~\nameref{sec:Discussion}.

\subsection*{Non-identical, dependent classifiers} 
\label{sec:nonid_dep} 

The most complex, but also the most realistic case, is when classifiers are non-identical and dependent. 
An analytical solution is not known \cite{Hisakado2006}, but simulations are straightforward. 
The results of \cite{Boland1989} can be extended, and we have that
\begin{align}
    P(Y_j|Y_0=1) &= \frac{\rho \sqrt{\theta_j (1-\theta_j) \theta_0 (1-\theta_0)} + \theta_j \theta_0 }{ \theta_0} \label{eq:dep_pos}\\
    \text{and} \nonumber\\ 
    P(Y_j|Y_0=0) &= \frac{-\rho \sqrt{\theta_j (1-\theta_j) \theta_0 (1-\theta_0)} + \theta_j (1-\theta_0) }{ (1-\theta_0)}. \label{eq:dep_neg}
\end{align}
Note that, since $P \leq 1$, it follows from Eq.~\ref{eq:dep_neg} that
\begin{align}
    \theta_j \geq \frac{\rho^2 \frac{\theta_0}{1-\theta_0}}{1+\rho^2 \frac{\theta_0}{1-\theta_0}}, \label{eq:rho_min}
\end{align}
and similarly for $0 \leq P$ and Eq.~\ref{eq:dep_pos}.

\subsection*{AUCs} 
\label{sec:AUCs}

In a binary classification problem, the calculation of the area under the receiver-operating characteristic curve (AUC) is based on scores of random variables from the positive and the negative class. 
The AUC is mathematically equivalent to the probability that a random variable from the positive class will be ranked higher, in terms of score, than a random variable from the negative class \cite{Hanley1982}.

Let $score(\cdot)$ be a function that assigns a score to a random variable. 
Let $S_+ = score(W_+)$, where $W_+$ is a random variable from the positive class, and $S_- = score(W_-)$, where $W_-$ is a random variable from the negative class.
The distribution of classifier scores, $S$, is a mixture distribution of the $S_+$ and the $S_-$ distributions, where the weight, $\pi$, corresponds to the probability of drawing a positive random variable, and has pdf
\begin{align*}
    \pi f_+(s) + (1-\pi)  f_-(s), 
\end{align*}
where $f_+(s)$ and $f_-(s)$ are the pdfs of the positive and the negative class, respectively. 

Then $\text{AUC} = P(S_- < S_+) = P(S_- - S_+ < 0)$, which is equivalent to the cdf at value $0$.
We use Normal distributions for illustration.
Let $S_+$ and $S_-$ have the distributions
\begin{align*}
    S_+ &\sim \mathcal{N}(\mu_+, \sigma^2_+) \\
    S_- &\sim \mathcal{N}(\mu_-, \sigma^2_-),
\end{align*}
where $\mathcal{N}(\mu, \sigma^2)$ is the Normal distribution with mean $\mu$ and variance $\sigma^2$.
We then have that
\begin{align}
    \text{AUC} = \Phi\left(\frac{0-(\mu_- - \mu_+)}{\sqrt{\sigma_-^2 + \sigma_+^2}} \right),
    \label{eq:AUC_normal}
\end{align}
where $\Phi$ is the cdf of the standard Normal distribution. 


For the moment, we have not included correlations in the simulations, but it should be possible to adopt a similar setup as in Sections \nameref{sec:id_dep} and \nameref{sec:nonid_dep}.
Let $S_{0^+}$ and $S_{0^-}$ be the scores of the reference classifier, for random variables from the positive and negative class, respectively. 

Let $\rho_+ = corr(S_{0^+},S_{j^+})$ be the correlation between the reference classifier scores $S_{0^+}$ and the scores of classifier $j$, $S_{j^+}$ for the positive class.
If they both follow Normal distributions as outlined above, then the bivariate distribution is  
$$
\begin{pmatrix}
S_{0^+}\\
S_{j^+}
\end{pmatrix}
\sim
\mathcal{N}\left(
\begin{pmatrix}
\mu_{0^+}\\
\mu_{j^+}
\end{pmatrix},
\begin{pmatrix}
\sigma_{0^+}^2 & \rho\sigma_{0^+}\sigma_{j^+} \\
\rho\sigma_{0^+}\sigma_{j^+} & \sigma_{j^+}^2 
\end{pmatrix}
\right),
$$
and equivalently for the negative class. 
However, the joint distribution of $(S_{0},S_{j})$ is not uniquely defined by the correlation, and additional assumptions must be made.

\subsection*{Estimating SOTA} 
\label{sec:estimate_SOTA}

When estimating $m \geq 3$ parameters simultaneously, the maximum likelihood estimator is inadmissible, meaning that an estimator with lower total mean squared error exists. 
This is known as Stein's paradox.
The James-Stein (JS) positive-part estimator is one such estimator, which is in itself inadmissible \cite{James1961Estimation}. 
However, Efron and Morris' 50 year old statement ``no uniformly better estimator has yet been found'' \cite{Efron1977Stein}, still holds.
Efron and Morris wrote a series of articles on the topic of the JS estimators and their generalizations. 
In \cite{Efron1973Stein}, they present a Bayesian derivation, and the empirical Bayes approach, where $\theta_j$ is a random variable estimated from the data. 


JS positive-part estimator is defined as follows for multivariate normal distributions with unknown mean vector, $\Theta$, and covariance matrix $\sigma^2I$ 
\begin{align}
    \Theta' = \Bigg ( 1 - \frac{(m-2)\hat{\sigma}^2}{||\hat{\Theta}(\mathbf{x})-\mathbf{w}||^2} \Bigg )_{+} \big(\hat{\Theta}(\mathbf{x}) - \mathbf{w}\big) + \mathbf{w},  \label{eq:JamesStein}
\end{align} 
where $\mathbf{w}$ is the shrink parameter and $()_+$ means that only positive elements contribute. 

The shrink parameter represents the Bayes' element: it is either chosen by the user (Bayes) or estimated from the data (empirical Bayes). 
An alternative is no weight, but $\mathbf{w} \neq 0$ has been shown to lower total mean squared error. 
Efron and Morris use the average in their examples. 
For large $m$, the ``data'' part of the average-weight estimator is close to zero, and it becomes optimal for identical classifiers.

For SOTA estimation, we want a distribution $\theta'_j \sim P(\theta')$ whose expected sample maximum equals the observed sample maximum, $E \max(\hat{\Theta}'(\mathbf{X})) = \max(\hat{\Theta})$.
We propose using JS positive-part estimator with the the $E \max(\hat{\Theta}'(\mathbf{X})) = \max(\hat{\Theta})$ restriction to estimate $\theta_{SOTA}$.
$\mathbf{w}$ is found numerically. 
We refer to it as the JS \textit{max} estimator. 


We demonstrate three versions of James-Stein; no weight, average and the proposed JS \textit{max}.


It is worth noting that James-Stein is not the only estimator among the group of shrink estimators. 
But not one that is guaranteed to be uniformly best. 

\section*{Simulations and real world examples} 
\label{sec:Simulations}

The following subsections exemplify and investigate different aspects of using the sample maximum as estimator for $\theta_{SOTA}$, in the context of public challenges.
Corresponding subsections in Section~\nameref{sec:Multiple} have details about the distributions. 


We use the following default parameter values: $n = 3,000$, $\theta=0.90$, $m=1,000$, $\theta_{SOTA} = 0.90$, $\theta_{0} = 0.90$, $AUC_{SOTA} = 0.90$.
In the hierarchical set-up, we use $B=1,000$ draws from $P(\theta)$ in Eq.~\ref{eq:hierarch_theta}, and $rep = 100,000$ when drawing random samples of $\mathbf{X}$ for each $\theta_j$. 
We have that 
$E \max \hat{\Theta}(\mathbf{X}) = E ( E ( \max \hat{\Theta}(\mathbf{X})|\Theta ) )$, 
and use Monte Carlo integration for calculations.  
We denote the estimates without the ``hat", 
leaning on the Law of Large Numbers and the access to arbitrary large $B$ and $rep$. 
We use significance level $\alpha = 0.05$ and two-sided CIs. 

The sampling procedure is illustrated in Fig.~\ref{fig:sampling_iid} for independent classifiers and in Fig.~\ref{fig:sampling_noniid} for dependent classifiers.

\begin{figure}[htb!]
    \centering
    \textbf{Sampling procedure for the independent model}\par\medskip  
    \begin{tikzpicture}[
      node distance=0.33cm and 1.5cm,
      every node/.style={align=center},
    ]
        \tikzset{
          sample/.style={-{Latex[length=3mm]}, thick}, 
          follow/.style={double equal sign distance, -{Stealth[open, length=3mm]}, thick} 
        }
        
        \node (prior) {\(P(\theta)\)};
        
        \node[above right = of prior] (theta1) {\(\theta_1\)};
        \node[below=of theta1] (theta2) {\(\theta_2\)};
        \node[below=of theta2] (dots) {\(\vdots\)};
        \node[below=of dots] (thetam) {\(\theta_m\)};
        
        \draw[sample] (prior) -- (theta1);
        \draw[sample] (prior) -- (theta2);
        \draw[sample] (prior) -- (thetam);
        
        \node[right=0.75cm of theta1] (px1) {\(P(X_1 \mid \theta_1)\)};
        \node[right=0.75cm of theta2] (px2) {\(P(X_2 \mid \theta_2)\)};
        \node[right=0.75cm of dots] (pxdots) {\(\vdots\)};
        \node[right=0.75cm of thetam] (pxm) {\(P(X_m \mid \theta_m)\)};
        
        \node[below left = 0.1cm and 0.75cm of thetam] (eq1b) {(\ref{eq:hierarch_theta})};
        
        \draw[follow] (theta1) -- (px1);
        \draw[follow] (theta2) -- (px2);
        \draw[follow] (thetam) -- (pxm);
        
        \node[right= of px1] (x1) {$x_1$};
        \node[right= of px2] (x2) {$x_2$};
        \node[below= of x2] (xdots) {\(\vdots\)};
        \node[right= of pxm] (xm) {$x_m$};
        \draw[sample] (px1) -- (x1);
        \draw[sample] (px2) -- (x2);
        \draw[sample] (pxm) -- (xm);
        
        \node[below left = 0.1cm and 0.75cm of xm] (eq1a) {(\ref{eq:hierarch_X})};
        
        \node[right=0.75cm of x1] (hat1) {$\hat\theta_1$};
        \node[right=0.75cm of x2] (hat2) {$\hat\theta_2$};
        \node[below= of hat2] (hatdots) {\(\vdots\)};
        \node[right=0.75cm of xm] (hatm) {$\hat\theta_m$};
        \draw[follow] (x1) -- (hat1);
        \draw[follow] (x2) -- (hat2);
        \draw[follow] (xm) -- (hatm);
        
        \node[below right = 0.75cm and 0cm of eq1a] (tail1) {};
        \node[right=1.5cm of tail1] (head1) {random sampling};
        \node[below = of tail1] (tail2) {};
        \node[right=0.75cm of tail2] (head2) {follows directly};
        \draw[sample] (tail1) -- (head1);
        \draw[follow] (tail2) -- (head2);

        \draw[decorate, decoration={brace}, thick]
          ($(hat1.north) + (1cm,0)$) -- ($(hat1 |- hatm.south) + (1cm,0)$)
          node[midway, right=4pt] {$\maxtheta$};
    \end{tikzpicture}
    \caption{The $\theta_j$'s are sampled from $P(\theta)$. The $x_j$'s are then sampled from their respective distributions, and $\hat{\theta}_{\max}$ follows directly.}
    \label{fig:sampling_iid}
\end{figure}

\begin{figure}[htb!]
    \centering
    \textbf{Sampling procedure for the dependent model}\par\medskip  
\begin{tikzpicture}[
    node distance=0.25cm and 1.5cm,
    every node/.style={align=center},
]
\tikzset{
    sample/.style={-{Latex[length=3mm]}, thick}, 
    follow/.style={double equal sign distance, -{Stealth[open, length=3mm]}, thick} 
}

\node[] (th0) {$\theta_0$};
\node[right = 0.75cm of th0] (py0) {$P(Y_0\mid\theta_0)$};
\node[right = of py0] (y0) {$\mathbf y_0 = \left\{ y_{01}, y_{02}, \ldots, y_{0n}\right\}$};
\draw[follow] (th0) -- (py0);
\draw[sample] (py0) -- (y0);

\node[below = 0.75cm of y0] (py1) {$P(Y_1\mid \mathbf y_0, \theta_1)$};
\node[below = of py1] (py2) {$P(Y_2\mid \mathbf y_0, \theta_2)$};
\node[below = of py2] (pydots) {$\vdots$};
\node[below = of pydots] (pym) {$P(Y_m\mid \mathbf y_0, \theta_m)$};

\draw[follow] (y0) -- (py1);

\node[left = 1cm of py1] (th1) {$\theta_1$};
\node[left = 1cm of py2] (th2) {$\theta_2$};
\node[below = of th2] (thdots) {$\vdots$};
\node[left = 1cm of pym] (thm) {$\theta_m$};

\draw[follow] (th1) -- (py1);
\draw[follow] (th2) -- (py2);
\draw[follow] (thm) -- (pym);

\node[right = of py1] (y1) {$\mathbf y_1$};
\node[right = of py2] (y2) {$\mathbf y_2$};
\node[below = of y2] (ydots) {$\vdots$};
\node[right = of pym] (ym) {$\mathbf y_m$};

\draw[sample] (py1) -- (y1);
\draw[sample] (py2) -- (y2);
\draw[sample] (pym) -- (ym);
 
\node[right = 0.75cm of y1] (x1) {$x_1$};
\node[right = 0.75cm of y2] (x2) {$x_2$};
\node[below = of x2] (xdots) {$\vdots$};
\node[right = 0.75cm of ym] (xm) {$x_m$};

\draw[follow] (y1) -- (x1);
\draw[follow] (y2) -- (x2);
\draw[follow] (ym) -- (xm);

\node[left = of th2] (pth) {$P(\theta)$};

\draw[sample] (pth) -- (th1);
\draw[sample] (pth) -- (th2);
\draw[sample] (pth) -- (thm);


\end{tikzpicture}
   \caption{The parameter $\theta_0$ defines the distribution from which the $Y_0$'s are sampled. The $\theta_j$'s are sampled from $P(\theta)$ as in Fig.~\ref{fig:sampling_iid}. Random samples of $Y_j$s are taken from their respective distributions, and $x_j$ follows directly, which in turn gives $\maxthetaX$ as in Fig.~\ref{fig:sampling_iid}. $P(Y_j|\mathbf y_0,\theta_j)$ is described in Eq.~\ref{eq:dep_pos} and \ref{eq:dep_neg}.}
    \label{fig:sampling_noniid}
\end{figure}

\begin{figure}[t!]
    \centering
    \textbf{Bias and standard deviation for non-iid classifiers}\par\medskip 
    \begin{subfigure}[t]{0.45\textwidth}
    \includegraphics[width=\linewidth]{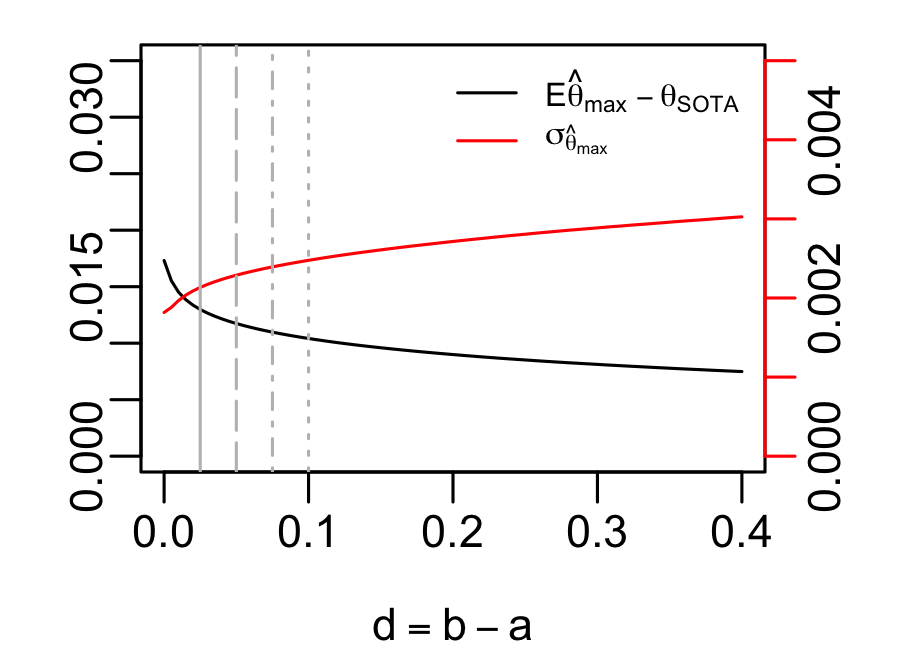}
    \caption{Bias and standard deviation as a function of $b-a$ in $\mathcal{U}(a,b)$} 
    \label{fig:bias_sd_d}
    \end{subfigure} 
    
    \begin{subfigure}[t]{0.45\textwidth}
    \centering
  \includegraphics[width=\linewidth]{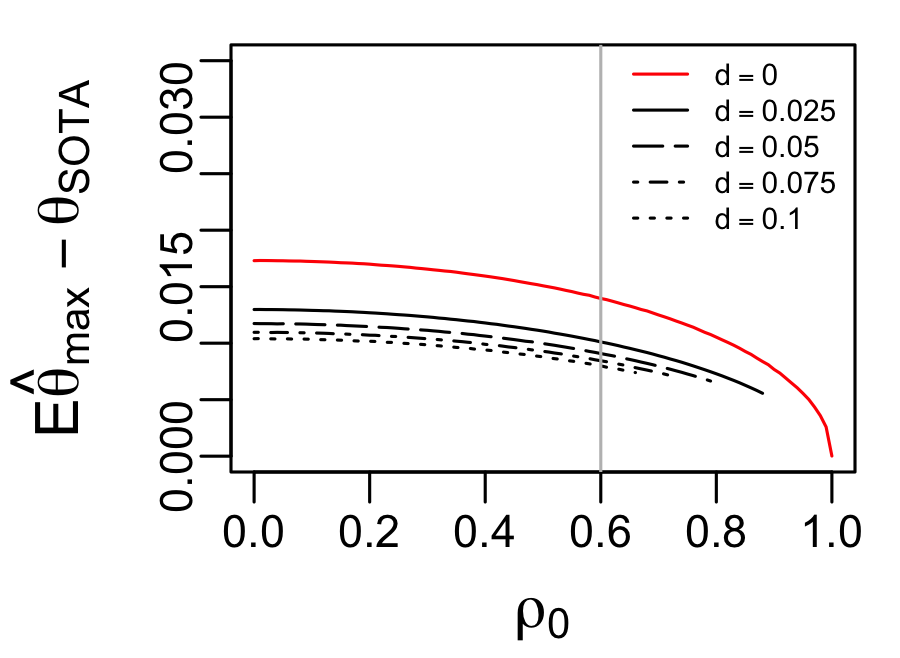}
    \caption{Bias as a function of correlation coefficient.}
    \label{fig:bias_thetamin_rho}
  \end{subfigure}
  \hfill
  \begin{subfigure}[t]{0.45\textwidth}
    \centering
  \includegraphics[width=\linewidth]{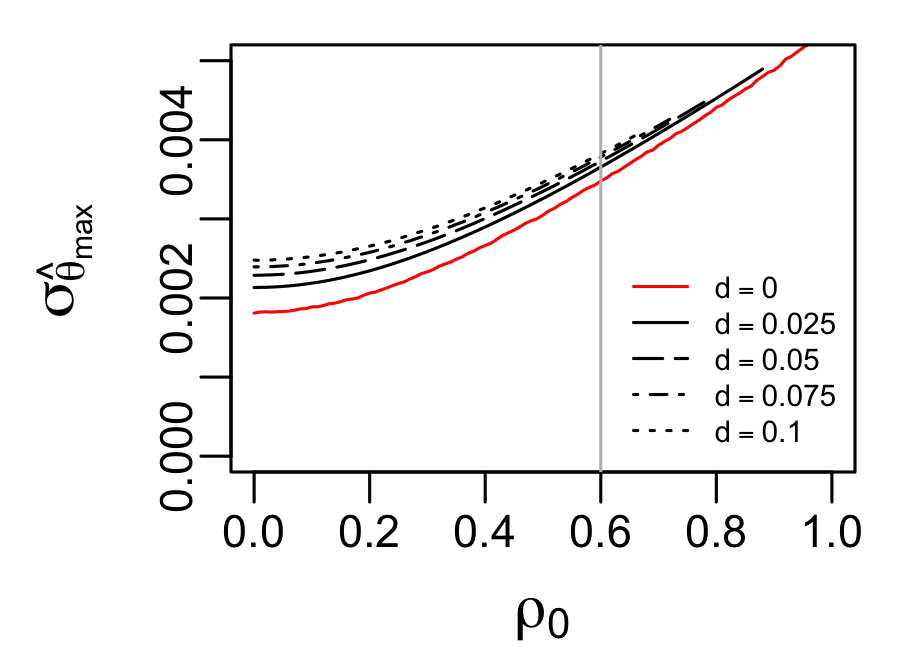}
    \caption{Standard deviation as a function of correlation coefficient.}
    \label{fig:sd_thetamin_rho}
  \end{subfigure}
  \caption{Bias and standard deviation for $\maxthetaX$ as a function of $d = b-a$ in $\mathcal{U}(a,b)$ for non-identical classifiers and of $\rho_0$ for conditional independent classifiers. 
  The scale of the $y$-axes are kept the same as in Fig.~\ref{fig:bias_sd_m_n_theta} for easier comparison. 
  (a) The left tip of the curve is identical classifiers. The support of $\mathcal{U}(a,b)$ stretches out along the x-axis, and the bias decreases. The intersection with the solid vertical line corresponds to the cdf and pmf in Fig.~\ref{fig:cdf_pmf_noniid_success}, and row 2 in Table~\ref{tab:noniid}. The intersections with the broken lines correspond to the left tips of the black lines in Fig.~\ref{fig:bias_thetamin_rho} where the classifiers are independent.  
  (b) The red line is identical classifiers, discussed in Section~\nameref{res:id_dep}. The left tip of the red curve corresponds to the iid~classifiers in Section~\nameref{res:id_indep}, and the intersection with the vertical line corresponds to row 3 in Table~\ref{tab:noniid}. The black lines show bias as a function of correlation coefficient demonstrated for different $d$'s. The intersection between the solid black curve and the grey vertical line corresponds to row 4 in Table~\ref{tab:noniid}. (c) The corresponding standard deviation. 
  }
  \label{fig:bias_sd_thetaminrho}
\end{figure}

The results are summarized in Table~\ref{tab:noniid} and Fig.~\ref{fig:bias_sd_thetaminrho}.
Table~\ref{tab:noniid} demonstrates that dependency and non-identical $\theta$s reduce the bias of the $\maxthetaX$ estimator, and assumptions regarding dependency and identicality are crucial to include in a model.
\begin{table}[ht!]
    \centering
    \textbf{Overview of $\maxthetaX$ for $\theta_{SOTA} = 0.90$} \par \medskip
    \begin{adjustbox}{max width=\textwidth}
    \begin{tabular}{ |c|c|c|c|c|c|c| } 
 \hline
  Section & Independent& Identical & Upper 95\% CI  & $\text{E}\maxtheta$ & $\sigma_{\maxtheta}$ \\ 
  \hline
\nameref{res:id_indep} &YES & YES & 0.9213 & 0.9173 & 0.0018 \\
  \hline
 \nameref{res:nonid_indep}& YES & NO & 0.9177 & 0.9129 & 0.0021 \\ 
 \hline
 \nameref{res:id_dep}& NO & YES & 0.9207 & 0.9140 & 0.0035 \\ 
 \hline
 \nameref{res:nonid_dep} & NO & NO & 0.9173 & 0.9101 & 0.0036 \\ 
 \hline
\end{tabular}
\end{adjustbox}
    \caption{The upper bound of the CI, the expected value and standard deviation for $\maxthetaX$ for different combinations of independency and identicality. The correlation coefficient and $P(\theta)$ are found in their respective subsections.}  
    \label{tab:noniid}
\end{table}
Fig.~\ref{fig:bias_sd_thetaminrho} shows how the bias and standard deviation change as functions of $\Theta$ and $\rho_0$ for these models, details are in the following sections. 

\subsection*{Identical, independent classifiers - simulations} 
\label{res:id_indep}

See Section~\nameref{sec:id_indep} for cdf and pmf. 

A different way to phrase the multiplicity problem is asking if a new and better classifier stands a chance up against the ensemble. 
Consider a public challenge with identical and independent classifiers, using the default parameter values from Section~\nameref{sec:Simulations}.
The classifier with the highest observed performance is declared winner.
Denote the CI of $\hat{\theta}_j = E\hat{\theta}_j = \theta_j$ by $(\theta_{\alpha/2},\theta_{1\text{-}\alpha/2})$. 
Consider a new classifier with probability of success $\theta_{m+1} = \theta_{1\text{-}\alpha/2}$.
By common scientific standards, an observed performance of $\hat{\theta}_{m+1} = \theta_{m+1}$ is considered significantly better than an observed performance of $\hat{\theta}_j = \theta_j$, but the multiplicity effect conceals this. 

\begin{figure}[htb!]
    \centering
    \textbf{The pmfs of two single $\hat{\theta}(X)$ compared to the $\maxtheta$ statistics}\par\medskip  
    \includegraphics[width=\linewidth]{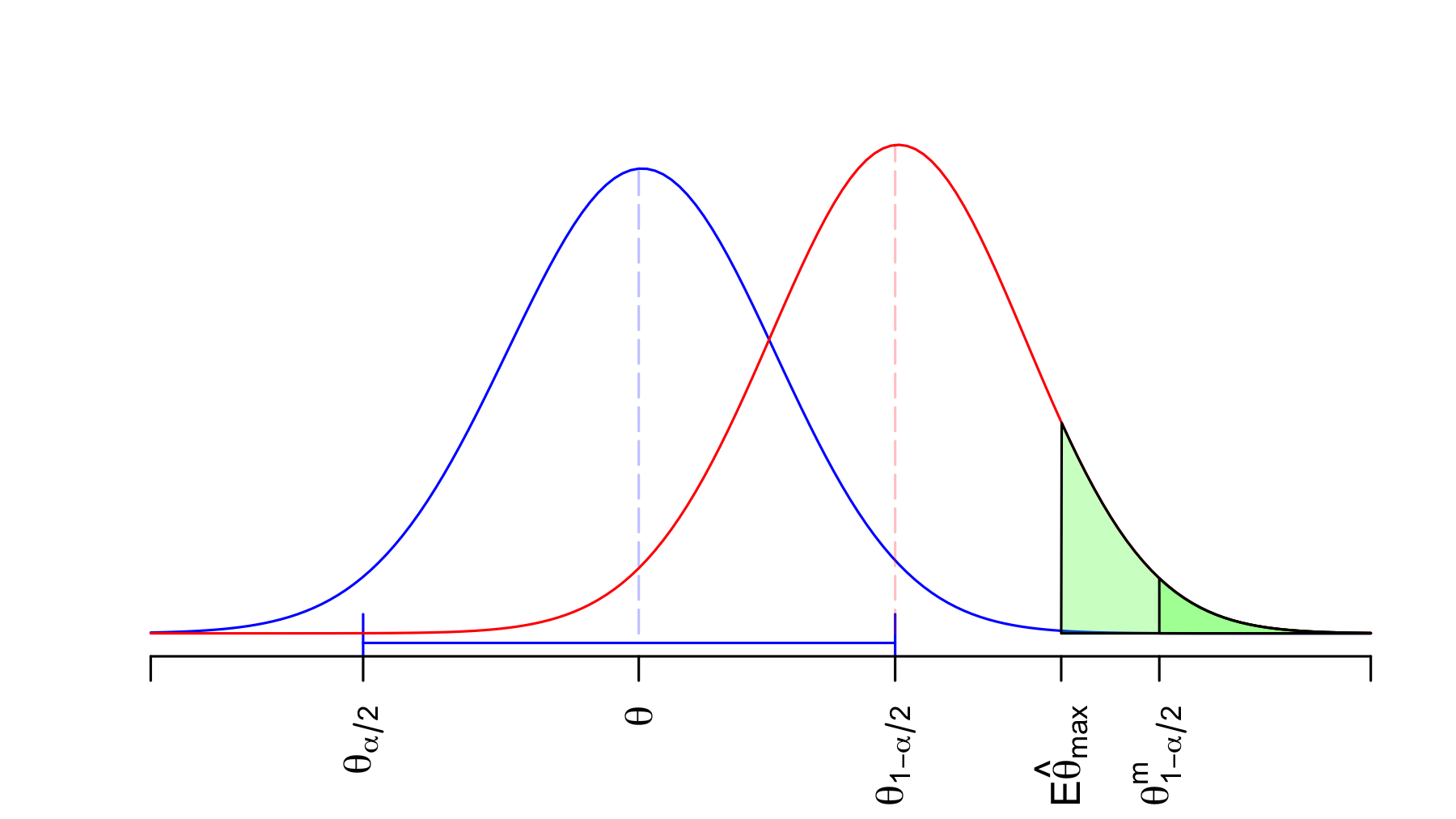}
    \caption{The blue and the red curves show the pmfs of $\hat{\theta}(X)$ for classifiers with probability of success $\theta$ and $\theta_{1\text{-}\alpha/2}$, respectively. 
    $\theta_{1\text{-}\alpha/2}$ is the upper limit of the CI of $\hat{\theta}(x) = \theta$. The CI is the blue horizontal line. 
    E$\maxtheta$ and $\theta^m_{1\text{-}\alpha/2}$ denote the expected value of $\maxtheta$ and the upper limit of the corresponding CI for $m$ classifiers. The green shaded areas are $P(\hat{\theta}_{m+1} \geq \theta^m_{1\text{-}\alpha/2})$ and $P(\hat{\theta}_{m+1} \geq$ \text{E}$\maxtheta)$.}
    \label{fig:multi_ci}
\end{figure}

In Fig.~\ref{fig:multi_ci}, the blue curve shows the pmf of a single $\hat{\theta}_j(X)$, 
and the CI of $\hat{\theta} = E\hat{\theta} = \theta$ is the blue line on the horizontal axis.
If we take multiplicity into account, the expected performance in terms of sample maximum for the $m$ participating teams, is $E\maxtheta$, indicated on the horizontal axis together with the upper limit of its CI, denoted as $\theta^m_{1\text{-}\alpha/2}$. 
For numerical values, see first row in Table~\ref{tab:noniid}.

The red curve shows the pmf of $\hat{\theta}_{m+1}(X)$.
The probability of the new and better classifier beating the ensemble of worse classifiers is low: $P(\hat{\theta}_{m+1} \geq \text{E}\maxtheta) = 0.10$, displayed as the green areas under the red curve. 
This means that by using the sample maximum as an estimator for SOTA, we can expect that 9 out of 10 classifiers that are significantly better will still be naively considered worse. 

\begin{figure}[t!]
    \centering
   \textbf{The bias and standard deviation of the sample maximum estimator} \\
    \begin{subfigure}[b]{0.45\textwidth}
    \centering
    \par\medskip  
  \includegraphics[width=\linewidth]{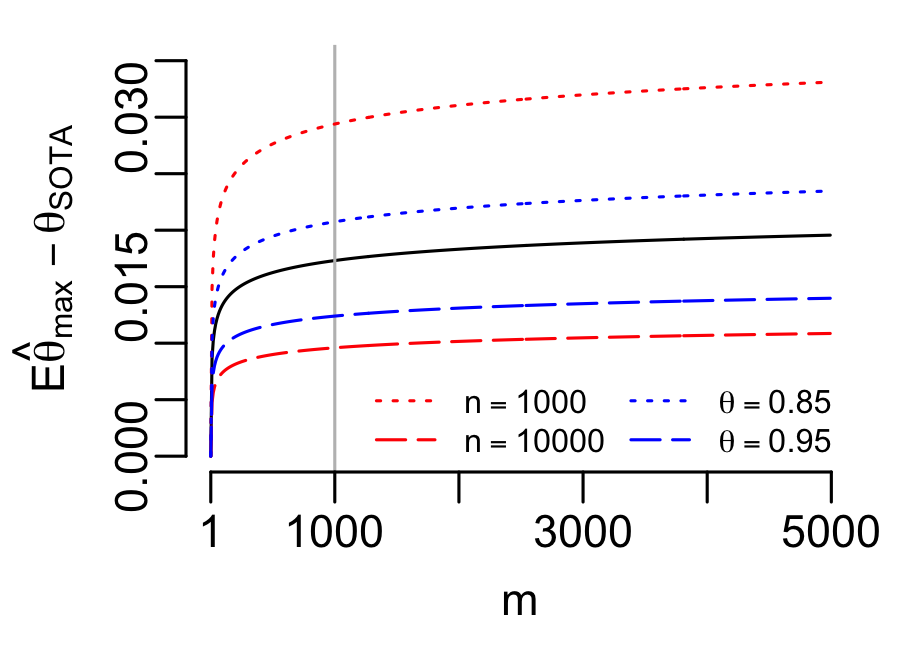}
  \caption{}
  \label{fig:bias_m}
  \end{subfigure}
  \hfill
  \begin{subfigure}[b]{0.45\textwidth}
    \centering
    \par\medskip  
  \includegraphics[width=\linewidth]{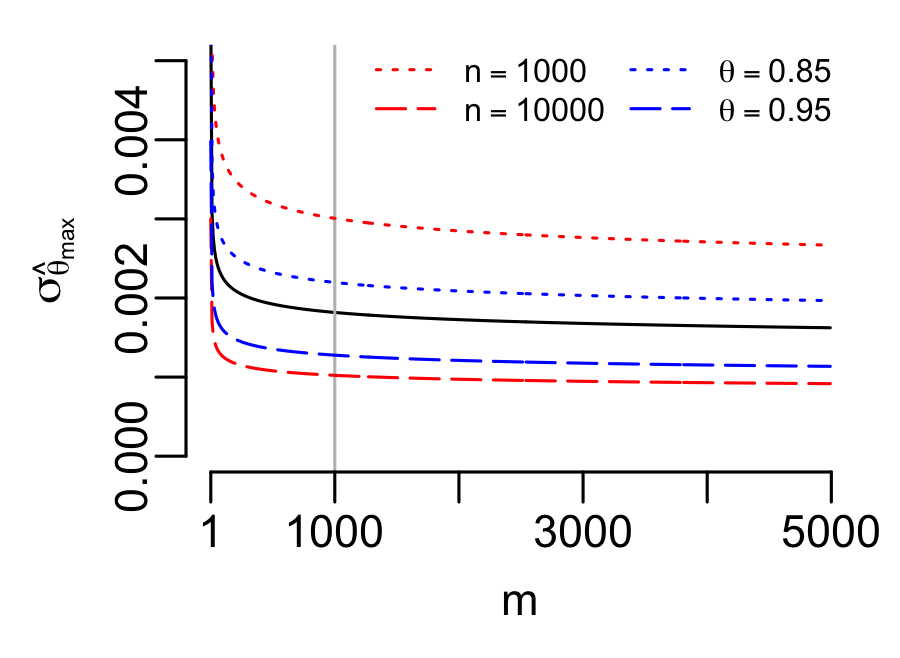}
  \caption{}
  \label{fig:sd_m}
  \end{subfigure}
    
  \begin{subfigure}[b]{0.45\textwidth}
    \includegraphics[width=\linewidth]{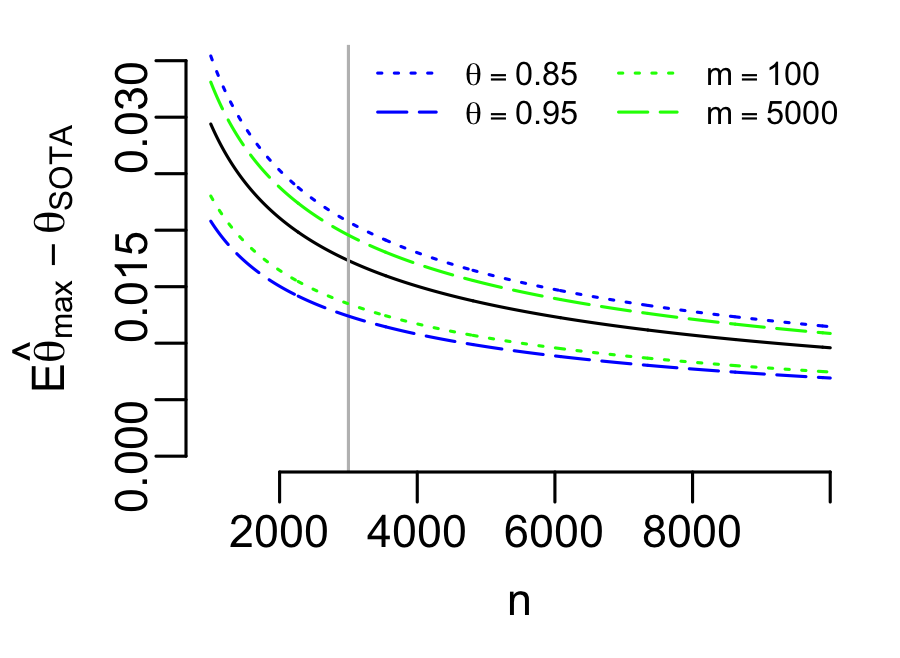}
    \caption{}
    \label{fig:bias_n}
    \end{subfigure}
    \hfill
  \begin{subfigure}[b]{0.45\textwidth}
    \includegraphics[width=\linewidth]{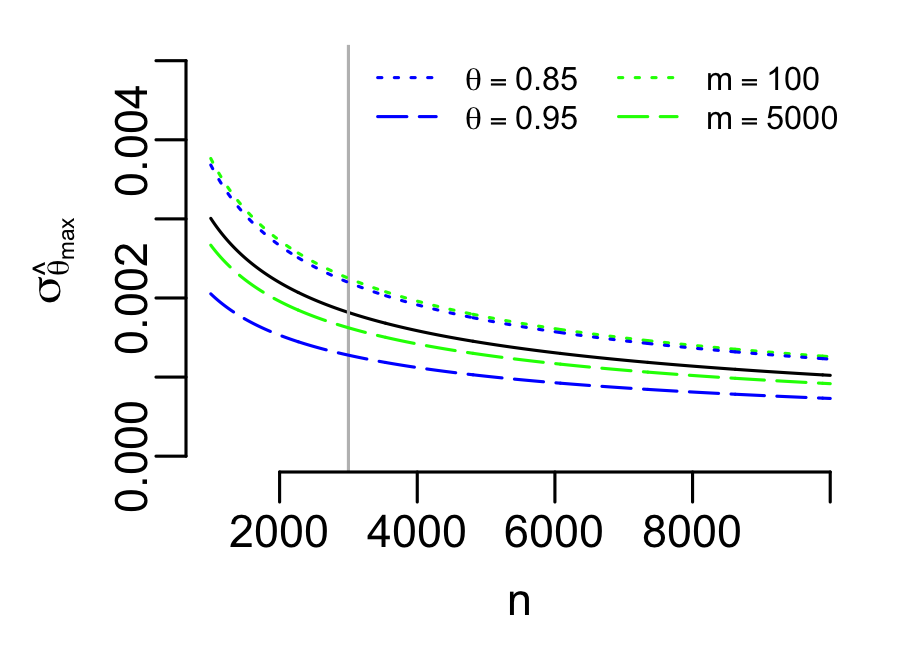}
    \caption{}
    \label{fig:sd_n}
    \end{subfigure}
  
  \begin{subfigure}[b]{0.45\textwidth}
    \centering
    \par\medskip  
  \includegraphics[width=\linewidth]{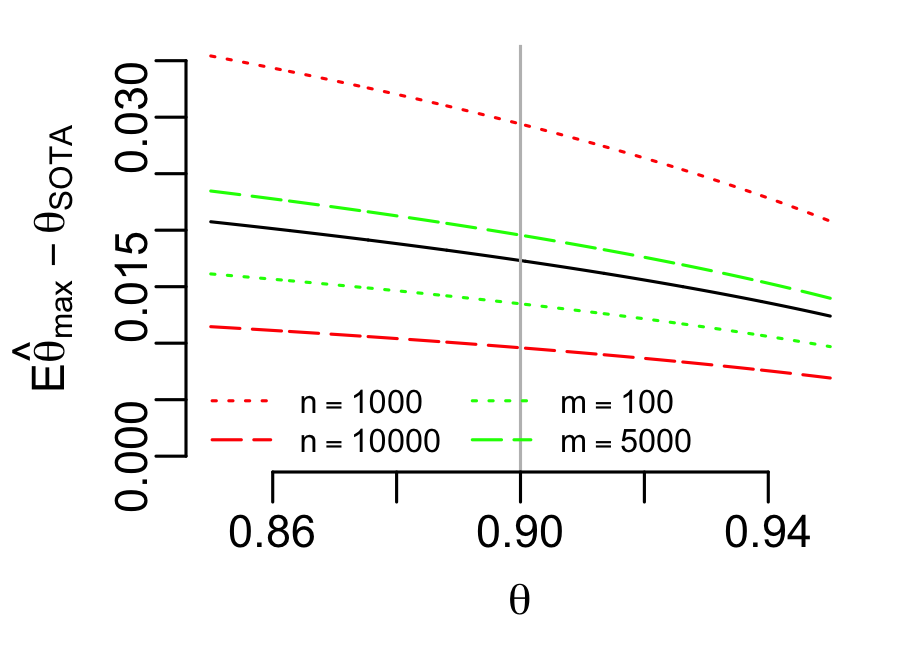}
  \caption{}
  \label{fig:bias_theta}
  \end{subfigure}
    \hfill
    \begin{subfigure}[b]{0.45\textwidth}
    \centering
    \par\medskip  
  \includegraphics[width=\linewidth]{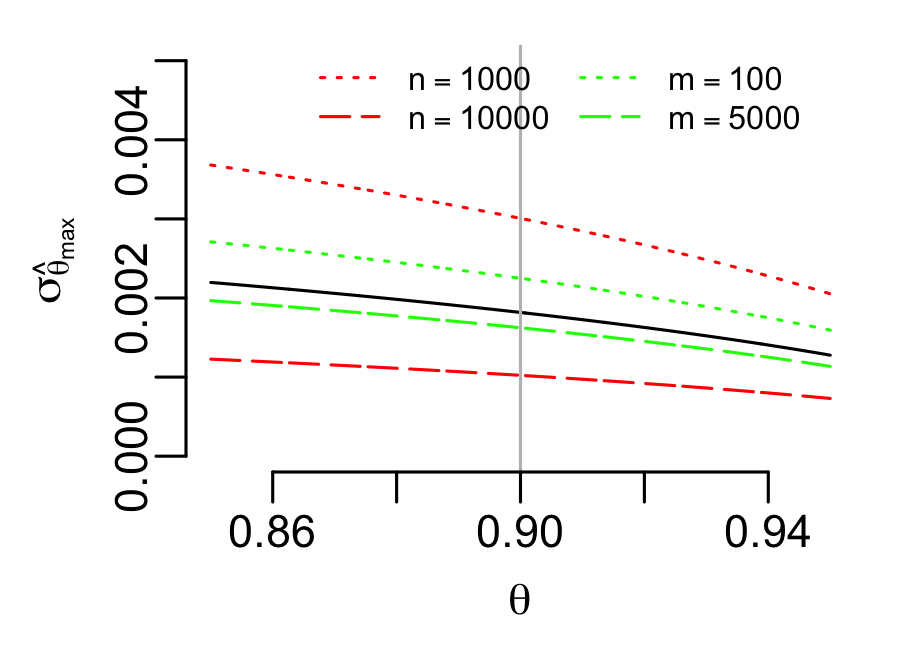}
  \caption{}
  \label{fig:sd_theta}
  \end{subfigure}
  \caption{Bias and standard deviation of $\maxthetaX$ as a function of each of its three parameters. For the black lines, two parameters are set to the default values, the third is on the x-axis. The other lines represent offsets. Intersections with the the gray vertical lines correspond to entries in Table~\ref{tab:mnp}.}
  \label{fig:bias_sd_m_n_theta}
\end{figure}

To get an insight in how the parameter values impacts the distribution, Fig.~\ref{fig:bias_sd_m_n_theta} shows the bias and standard deviation of $\maxthetaX$ as a function of $m$, $n$ and $\theta$, respectively. 
We see that the impact of number of classifiers is dramatic, and multiplicity considerations must be treated with full attention starting at $m=2$.
Fig.~\ref{fig:bias_n} demonstrates that the test set size also has large impact on the bias, which can be traced back to the standard error, $\sqrt{\theta(1-\theta)/ n}$. 
The standard error also explains the decrease in bias as $\theta$ approaches 1, as seen in Fig.~\ref{fig:bias_theta}: The standard deviation of the individual $\hat{\theta}_j(X)$'s is at its highest for $\theta = 0.5$ and decreases monotonically. 

Figs.~\ref{fig:sd_m}, \ref{fig:sd_n} and \ref{fig:sd_theta} show the corresponding standard deviations. 
For numerical values, see Table~\ref{tab:mnp}.

\begin{table}[ht!]
    \centering
    \textbf{Expected values and standard deviations for $\maxthetaX$}
    \begin{adjustbox}{max width=\textwidth}
    \par \medskip
    \begin{tabular}{ |c|c|c|c|c|c| } 
 \hline
  m& n & $\theta$ & E$\maxtheta$ & $\sigma_{\maxtheta}$ \\ 
  \hline
  \textbf{1,000} & \textbf{3,000} & \textbf{0.90} & \textbf{0.9173} & \textbf{0.001817} \\ 
 \hline
{\color{green}100} & {3,000} & {0.90} & 0.9135 & 0.002250 \\ 
  \hline
 {\color{green}5,000} & 3,000 & 0.90 & 0.9196 & 0.001623 \\ 
 \hline
 1,000 & {\color{red}1,000} & 0.90 & 0.9294 & 0.003007 \\
 \hline
 1,000 & {\color{red}10,000} & 0.90 & 0.9096 & 0.001022 \\
 \hline
 {1,000} & {3,000} & {\color{blue}0.85} & 0.8707 & 0.002197 \\
    \hline
    {1,000} & {3,000} & {\color{blue}0.95} & 0.9624 & 0.001277 \\
 \hline
\end{tabular}
\end{adjustbox}
    \caption{The expected value and standard deviation for the SOTA estimator with varying test set size, number of classifiers and probability of success. Row 1 is the example used in Fig.~\ref{fig:cdf_pmf_success}. The subsequent rows correspond to the intersection between the gray vertical lines and the curves in Fig.~\ref{fig:bias_sd_m_n_theta}.}
    \label{tab:mnp}
\end{table}

\subsection*{Non-identical, independent classifiers - simulations} 
\label{res:nonid_indep}

See Section~\nameref{sec:nonid_indep} for cdf and pmf. 

From row 2 in Table~\ref{tab:noniid} we see that the bias is smaller, but the standard deviation is larger compared to the iid case, 
as the classifiers at the lower end do not contribute to the sample maximum, and the reduced number of contributing classifiers gives a higher variance. 

Fig.~\ref{fig:bias_sd_d} displays the bias and standard deviation as a function of $b-a$ in $\mathcal{U}(a,b)$. 
The bias decreases rapidly at first, but the rate goes down fast.
This means that taking into account that classifier performances are heterogeneous is important, and even small variations from identicality can have great impact on SOTA estimation.

\subsection*{Identical, dependent classifiers - simulations} 
\label{res:id_dep}
See Section~\nameref{sec:id_dep} for details. 

We give an example for the case where $\rho_{0j} = \rho_{0}$ and $\theta_j = \theta$ for all classifiers, setting $\rho_{0}$ based on the work of \cite{Mania2019},
who calculated model similarity, defined as the probability of two classifiers giving the same output, which is equivalent to 
\begin{align*}
    (1-\theta_j)(1-\theta_k) + \theta_j\theta_k + 2\rho_{jk} \sqrt{\theta_j\theta_k(1-\theta_j)(1-\theta_k)}
\end{align*} 
for models with correlation $\rho_{jk}$.
The average accuracy for top-performing ImageNet models, $\hat{\theta} = 0.756$, and the calculated similarity of $0.85$ gives $\rho = 0.36$ under the assumption of equal correlation for all classifiers.
We have that $\rho_{jk} = \rho_{0j}^2$, and hence our proposed correlation $\rho_{0}$ is set to $0.6$. 

Comparing rows 1 and 3 in Table~\ref{tab:noniid}, we see a decrease in bias compared to the iid classifiers. 
However, the standard deviation has increased due to the random nature of $Y_0$, so the upper CI is essentially the same. 

To give an idea of how the randomness of $Y_0$ influences the standard deviation, we fix the $y_0$'s with exactly $\theta_0 = 0.90$ correct predictions.
The standard deviation decreases from $0.0035$ to $0.0015$.
For real datasets, the dependencies have both random and fixed sources. 
The training and the test samples are random sources, a fixed source could be pre-trained models. 

For the whole range of $\rho_{0j}$, see the red curve in Fig.~\ref{fig:bias_thetamin_rho}.
We see here that the correlation coefficient value in a model has a great impact on the estimated bias if the correlation is substantial. This can explain that in mature fields, where most competing teams use variants of the same model, the multiplicity will not impact the overestimation of SOTA as much as in young fields where the competing teams have more diverse models. This aligns with what is often observed: overoptimistic results in the beginning, and more sober as the field evolves.

The bias decreases slowly for low correlations, but the rate increases as $\rho_0$ increases. 
Note that Eq.~\ref{eq:rho_min} restricts $d = b-a$, and the graphs are right-truncated.
We observe in Fig.~\ref{fig:bias_sd_d} that change in $d$ has less impact on the higher end of the scale, which supports a decision of excluding teams with low performance.

\subsection*{Non-identical, dependent classifiers - simulations} 
\label{res:nonid_dep} 
See Section~\nameref{sec:nonid_dep} for details. 

The results in Table~\ref{tab:noniid} show that this non-identical, dependent model has the lowest bias, but the highest standard deviation. 
This sums of up much of Table~\ref{tab:noniid}, where more complex, and therefore more realistic, models reduces the bias but increases the variances. 

Fig.~\ref{fig:bias_thetamin_rho} shows the impact of correlation for non-identical classifiers. 

\subsection*{JS estimators - simulations}
\label{res:JS}

We here demonstrate the use of JS positive part estimator, Eq.~\ref{eq:JamesStein}, applied to the four simulation schemes described in detail above. 
The assumptions for Eq.~\ref{eq:JamesStein}, are fulfilled only for identical, independent classifiers (normal approximation of binomial distribution). 
Generalizations into non-identical variances and non-normality exist \cite{Efron1973Stein}, but the gain of using more complex estimators is questionable when the underlying distribution is unknown. 

The results are displayed in Table~\ref{tab:js}, together with
column four from Table~\ref{tab:noniid} for reference, referred to as Maximum Likelihood Estimator (MLE).

  \begin{table}[ht!]
    \centering
    \textbf{Estimating $\theta_{SOTA}$}\\
    \begin{adjustbox}{max width=\textwidth}
    \par \medskip
    \begin{tabular}{ |l|c|c|c|c|} 
 \hline
  Simulation setup & $\mathbf w = \mathbf 0$ & average &  JS \textit{max} & MLE\\ 
  \hline
  Identical, indep. & 0.9174 & 0.9000     &    0.9125     & 0.9173\\
  Non-identical, indep. & 0.9129 & 0.8875 &    0.9084     & 0.9129\\
  Identical, dep. & 0.9138 & 0.8999 &          0.9077     & 0.9140\\
  Non-identical, dep. & 0.9098 & 0.8874 &      0.9048     & 0.9109\\
 \hline
\end{tabular}
\end{adjustbox}
    \caption{$\theta_{SOTA}$ estimates for the simulations in Sections~\nameref{res:id_indep}, \nameref{res:nonid_indep}, \nameref{res:id_dep} and \nameref{res:nonid_dep}.}
    \label{tab:js}
\end{table}


As expected, the average weight performs optimal for identical distribution of $\theta$, and the unweighted estimator performs the worst overall.
The $E \max(\hat{\Theta}'(\mathbf{X})) = \max(\hat{\Theta})$ weight overestimates $\theta_{SOTA}$.

\subsection*{AUCs - simulations} 
\label{res:AUCs}
See Section~\nameref{sec:AUCs} for details. 

\begin{figure}[t!]
     \centering
     {\bf Sample maximum AUC}\\
     \begin{subfigure}[t]{0.45\textwidth}
         \centering
        \includegraphics[width=\textwidth]{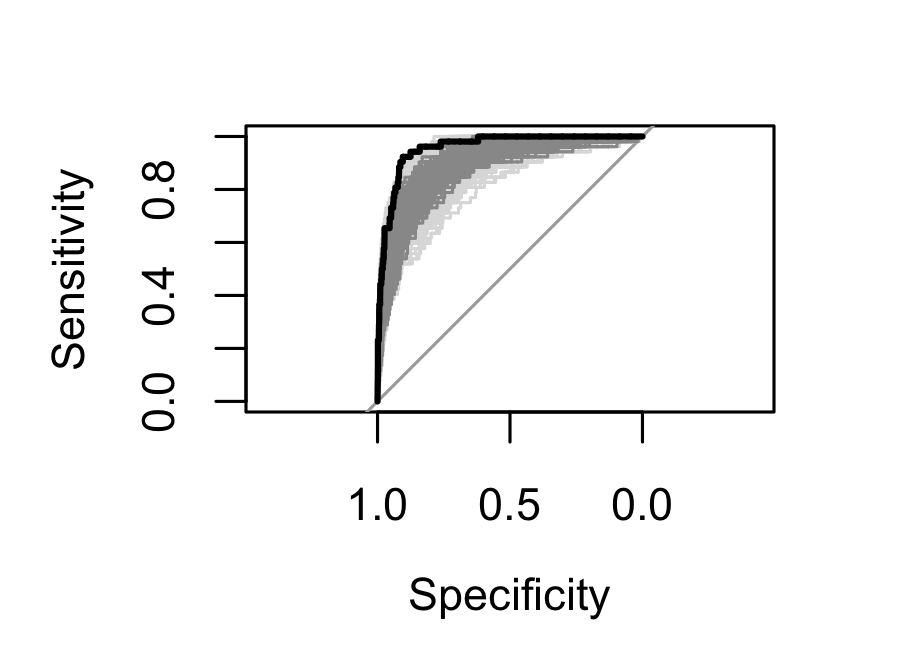}
         \caption{ROC curves. The curves that fall within the inner quartile of all AUCs are dark grey, the ones outside are light grey. The black line is the sample maximum AUC of 0.9580.}
         \label{fig:roc_simu}
     \end{subfigure}
    \begin{subfigure}[t]{0.45\textwidth}
    \centering
    \includegraphics[width=\textwidth]{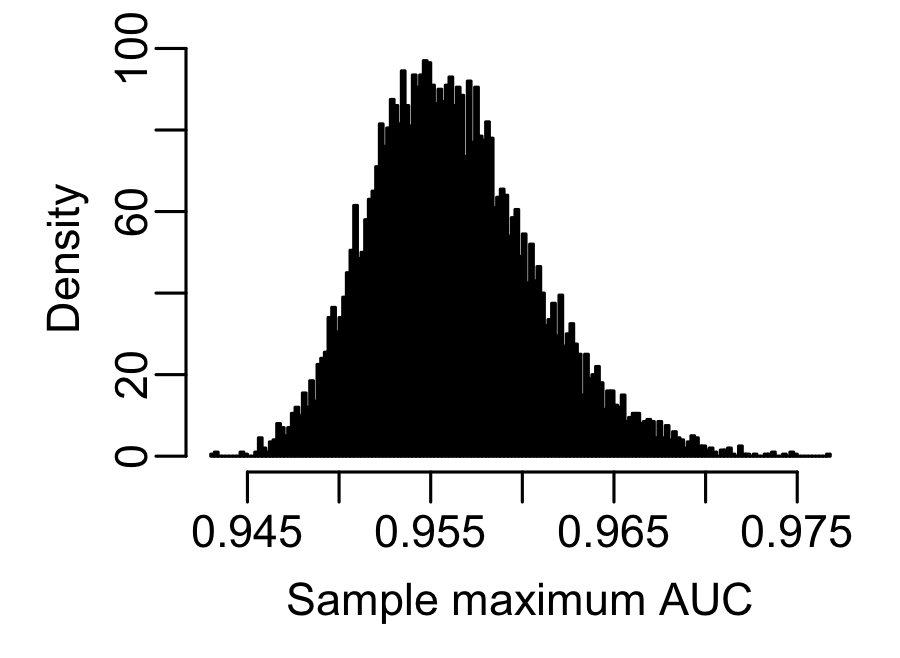}
    \caption{A simulated distribution of the sample maximum AUC. This is analogous figure to {Fig. \ref{fig:pmf_success}} in Section~\nameref{sec:id_indep}.}
    \label{fig:auc_09}
\end{subfigure}

     \caption{Simulated example for identical classifiers with $AUC = 0.90$ and class balance $\pi = 0.017$}
    \label{fig:roc_gallery}
\end{figure}

We give an example for the same class balance as in the Melanoma challenge (Section~\nameref{sec:Melanoma}), $\pi = 0.017$, as this reflects real world examples. 
The classifiers are independent, with identical $AUC=0.90$.
Fig~\ref{fig:roc_simu} shows ROC curves, 
where the sample maximum AUC is 
displayed as a black line. 
Fig.~\ref{fig:auc_09} shows a simulated distribution of the sample maximum. 

The simulated expected value is $0.9562$, 
the standard deviation is $0.004459$, 
and the simulated 95\% CI is $(0.9486, 0.9662)$. 
Comparing this to the results in the upper row of Table~\ref{tab:mnp}, we see that the bias is much larger. 
This is partly due to the large variance created by the class imbalance. 
Although the sample size is $n=3,000$, only $51$ of those belong to the positive class, creating large variance in the estimator.

\subsection*{Estimating SOTA - real world examples} \label{res:estimate_SOTA}

We here demonstrate different aspects of real world analysis through SOTA estimation of 
Obesity Risk, 
Cassava Leaf 
and Melanoma. 
Obesity Risk and Cassava Leaf have accuracy, defined as $\hat{\theta}$, as ranking criterion, whereas Melanoma uses the more common AUC.
Table~\ref{tab:jsReal} sums up the results, but there is no ground truth, as in Table~\ref{tab:js}.

\begin{table}[ht!]
    \centering
    \textbf{James-Stein estimate for real world examples}
    \begin{adjustbox}{max width=\textwidth}
    \par \medskip
    \begin{tabular}{ |l|c|c|c|c| } 
 \hline
  Dataset & $\mathbf w = \mathbf 0$ & average & JS \textit{max}   & MLE\\ 
  \hline
  Obesity risk & 0.8888 & 0.8687 & 0.9050 & 0.9116\\
  Cassava & 0.8749 & 0.8212 & 0.8943 & 0.9132\\
  Melanoma & 0.9399 & 0.8798 &  0.8991 & 0.9491\\
 \hline
\end{tabular}
\end{adjustbox}
    \caption{JS positive-part estimates for the challenges in  Sections~\nameref{sec:ObesityRisk}, \nameref{res:Cassava}, and \nameref{sec:Melanoma}.}
    \label{tab:jsReal}
\end{table}

The JS \textit{max} estimator gives higher estimates than the JS average estimator, but they are not guaranteed to be overestimations. 
The JS estimators are not suited for the Cassava data set, as discussed in Section~\nameref{res:Cassava}, but are presented here for the interested reader.

\subsubsection*{Multi-Class Prediction of Obesity Risk} 
\label{sec:ObesityRisk}

\begin{figure}[t!]
     \centering
     {\bf Obesity risk: $\hat\Theta$, $\Theta'$ and corresponding realisations}\\
     \begin{subfigure}[t]{0.45\textwidth}
         \centering
         \includegraphics[width=\textwidth]{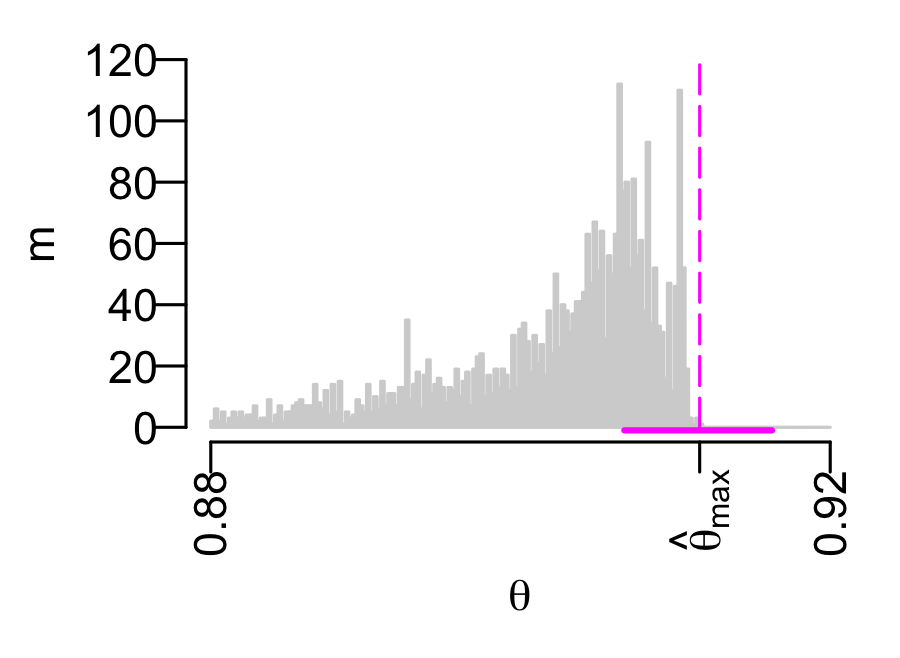}
         \caption{The observed performances, $\hat \Theta$. $\hat{\theta}_{\max}$ and its CI in magenta. For exact numbers, see row 1, Table~\ref{tab:obesity}}.
         \label{fig:obesity_kaggle}
     \end{subfigure}
     \hfill
     \begin{subfigure}[t]{0.45\textwidth}
         \centering
        \includegraphics[width=\textwidth]{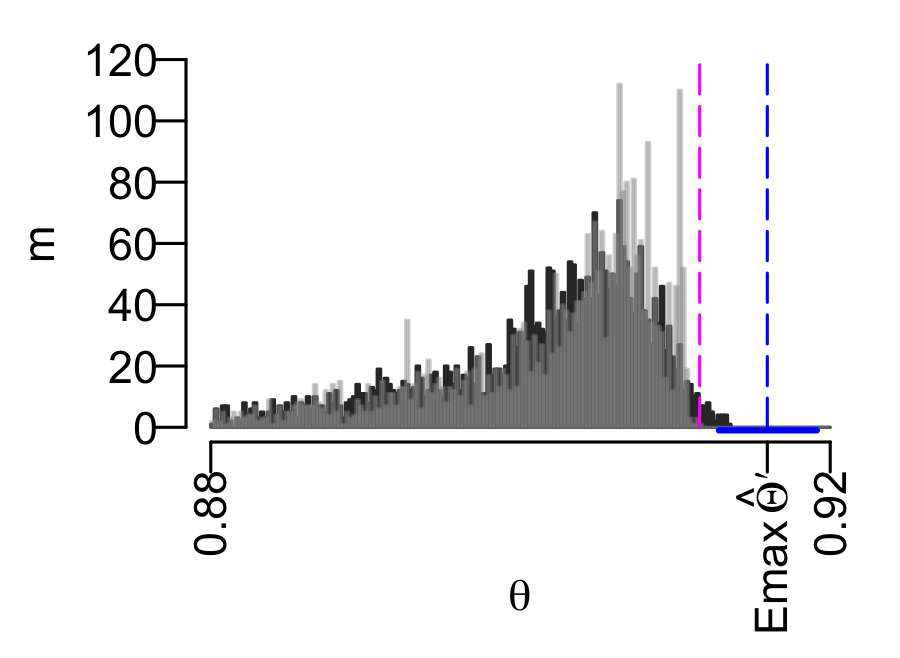}
         \caption{A realisation from a), $\text{E} \max (\hat{\Theta}')$ and its CI in blue. For exact numbers, see row 2, Table~\ref{tab:obesity}.}
         \label{fig:obesity_direct_sampling}
     \end{subfigure}
     \begin{subfigure}[t]{0.45\textwidth}
         \centering
         \includegraphics[width=\textwidth]{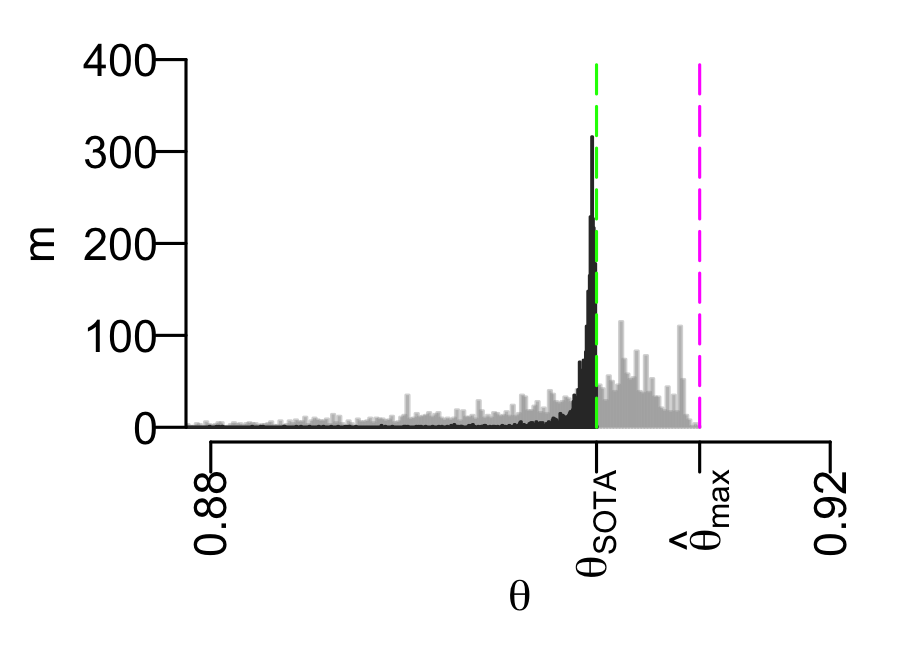}
         \caption{The JS \textit{max}-shrunk data, $\Theta'$ (black) and $\hat \Theta$ (gray) for reference. The new estimated $\theta_{SOTA}$ (green) and the MLE (magenta) for reference.}
         \label{fig:obesity_shrunk_for_expect}
     \end{subfigure}
     \hfill
     \begin{subfigure}[t]{0.45\textwidth}
         \centering
        \includegraphics[width=\textwidth]{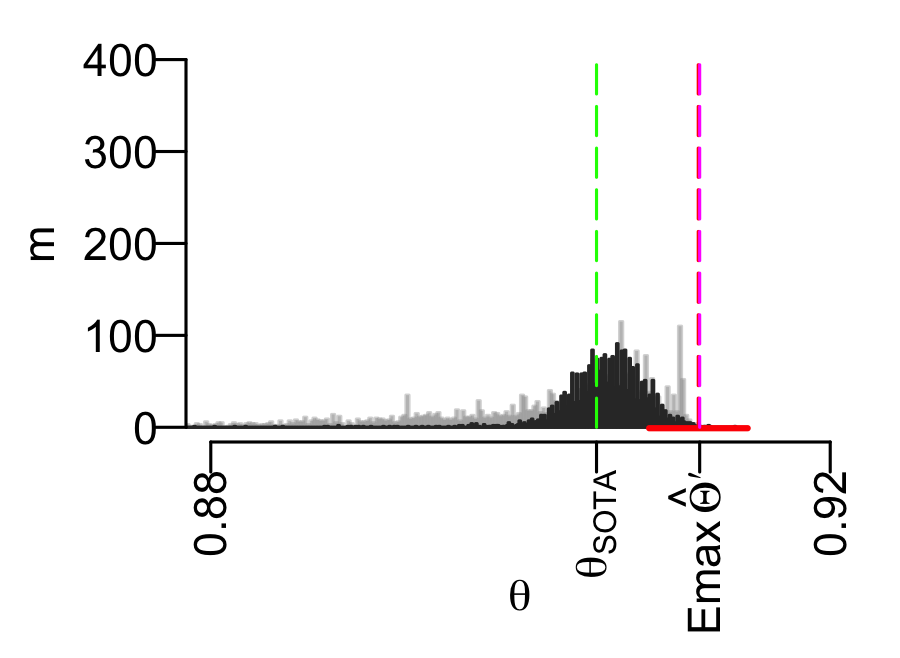}
        \caption{A realisation from c), $\text{E} \max (\hat{\Theta}') =\hat{\theta}_{\max}$ and its CI (red). For exact numbers, see row 3, Table~\ref{tab:obesity}.}
         \label{fig:obesity_shrunk_for_expect_realisation}
     \end{subfigure}
     \caption{The upper row shows the observed performances and a corresponding realisation. The lower row shows a shrunk version and a corresponding realisation.}
    \label{fig:obesity}
\end{figure}

\textit{Multi-Class Prediction of Obesity Risk} is a challenge in the 2024 Kaggle Playground Series, using a synthetic data set generated from the \textit{Obesity or CVD risk} data set \cite{Palechor2023}. 
The test sample size was $n=13,840$, and the number of teams were $m = 3,558$, excluding teams with accuracy lower than chance. 
Fig.~\ref{fig:obesity_kaggle} shows the observed performances, $\hat{\theta}_j$'s, zoomed in at the 
those above $0.88$ for better visualisation.
The observed sample maximum is $\maxtheta = 0.9116$, 
shown as the vertical dotted magenta line in all subfigures together with the $\hat{\theta}_j$'s in gray, for reference. 

The CI for $\maxtheta$ is shown as the magenta horizontal line in Fig.~\ref{fig:obesity_kaggle}, enveloping $921$ teams. 
This should raise suspicion that the right tail seen 
can be the product of multiplicity. 
For exact numbers, see Table~\ref{tab:obesity}, where the entries are colour coded to match the figures. 

We demonstrate the impact of ignoring the multiplicity effect by setting the proposed  distribution ${\theta}'_1, \ldots, {\theta}'_m$ equal to the observed performances, $\hat{\theta}_j$'s, and calculate the expected value of the sample maximum, $\text{E} \max (\hat{\Theta}')$. 
We use the set-up from Section~\nameref{sec:nonid_dep}. 
A realization is shown in Fig.~\ref{fig:obesity_direct_sampling}.
The upper tail is stretched according to the variances of the $Y_j$s.
The expected sample maximum, $\text{E}\max (\hat{{\Theta}}') = 0.9159$, shown as the vertical dotted blue line, is far above the observed sample maximum in the challenge. 
The lower bound of its CI, shown as the blue horizontal line in Fig.~\ref{fig:obesity_direct_sampling}, exceeds $\max (\hat{\Theta})$, and we can conclude that the observed sample maximum is an unlikely candidate for SOTA.

Fig.~\ref{fig:obesity_shrunk_for_expect} shows the proposed shrunk distribution, $\Theta'$.
Fig.~\ref{fig:obesity_shrunk_for_expect_realisation} shows one realization from  $\Theta'$, where $\max (\hat{\Theta}')$ coincides with $\max(\hat{\Theta})$, and its CI.
$\max ({\Theta}') = 0.9050$ and this serves as a new candidate for the SOTA. 


\begin{table}[ht!]
    \centering
    \textbf{Estimating ${\theta}_{SOTA}$ for Obesity risk} \par \medskip
    \begin{adjustbox}{max width=\textwidth}
    \begin{tabular}{ |c|c|c|c|} 
 \hline
  Approach & $\max (\Theta')$ & 95\% CI  & $\text{E}\max (\hat{\Theta}')$ 
  \\ 
  \hline
  \hline
  no mult.  & {\color{magenta}$\maxtheta$} & {\color{magenta}(0.9067, 0.9162)} & - \\ 
  \hline
$\Theta' = \hat \Theta$  & {\color{magenta}$\maxtheta$} & {\color{blue}(0.9128, 0.9191)} & {\color{blue}0.9159} \\ 
  \hline 
 JS \textit{max}  & {\color{green}0.9050} & {\color{red}(0.9084, 0.9148)} &  {\color{magenta}$\maxtheta$} 
 \\
  \hline
\end{tabular}
\end{adjustbox}
    \caption{The 95\% CI and the expected value for sample maximum, $\max (\hat{\Theta}')$, for different SOTA estimation strategies, where SOTA is defined as $\max (\Theta')$. The challenge sample maximum is $\maxtheta= 0.9116$. 
    }
    \label{tab:obesity}
\end{table}

Table~\ref{tab:obesity} demonstrates that taking the effect of multiplicity into account changes how results are perceived. 
The new SOTA value is $0.9050$, and even though its reduction from best performance is only $0.0066$, the effect it has on who we consider performing at SOTA-level is dramatic: around a quarter of the classifiers have sample performance above the new SOTA. 

\subsubsection*{Cassava Leaf Disease} 
\label{res:Cassava}

In this example, the observed sample maximum is unlikely to be the product of multiplicity. 
In \textit{The Cassava Leaf Disease Classification} challenge \cite{Mwebaze2020}, the training sample consisted of $21,367$ images of five classes, and the test sample of approximately $15,000$ images. 
$3,752$ teams participated, excluding those with sample performance lower than chance. 

\begin{figure}[t!]
     \centering
     {\bf Cassava Leaf: $\hat \Theta$ and corresponding realisations}\\
     \begin{subfigure}[t]{0.45\textwidth}
         \centering
         \includegraphics[width=\textwidth]{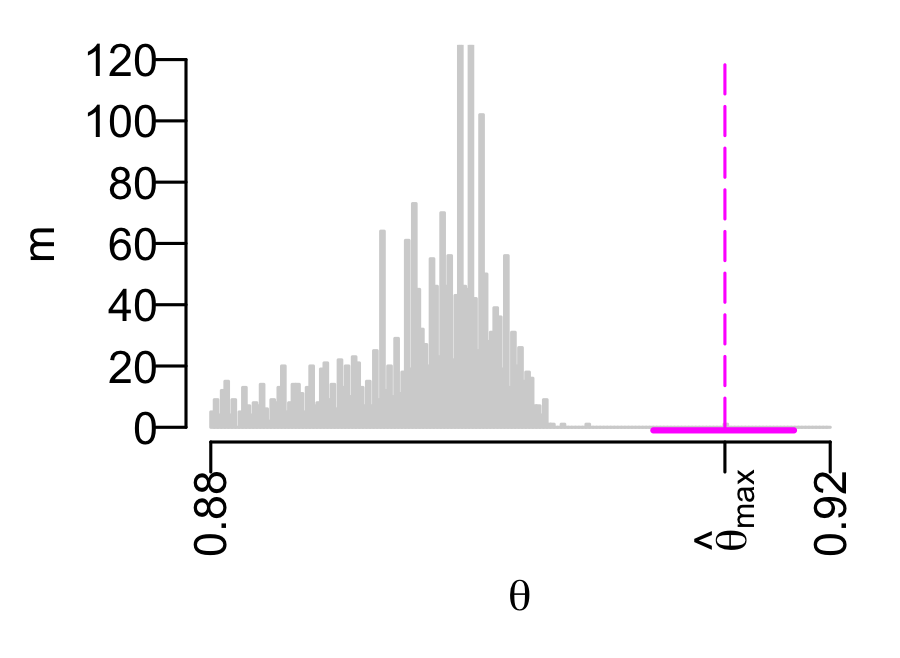}
         \caption{The observed performances, $\hat \Theta$. $\hat{\theta}_{\max}$ and its CI.}
         \label{fig:casava_kaggle}
     \end{subfigure}
     \hfill
     \begin{subfigure}[t]{0.45\textwidth}
         \centering
         \includegraphics[width=\textwidth]{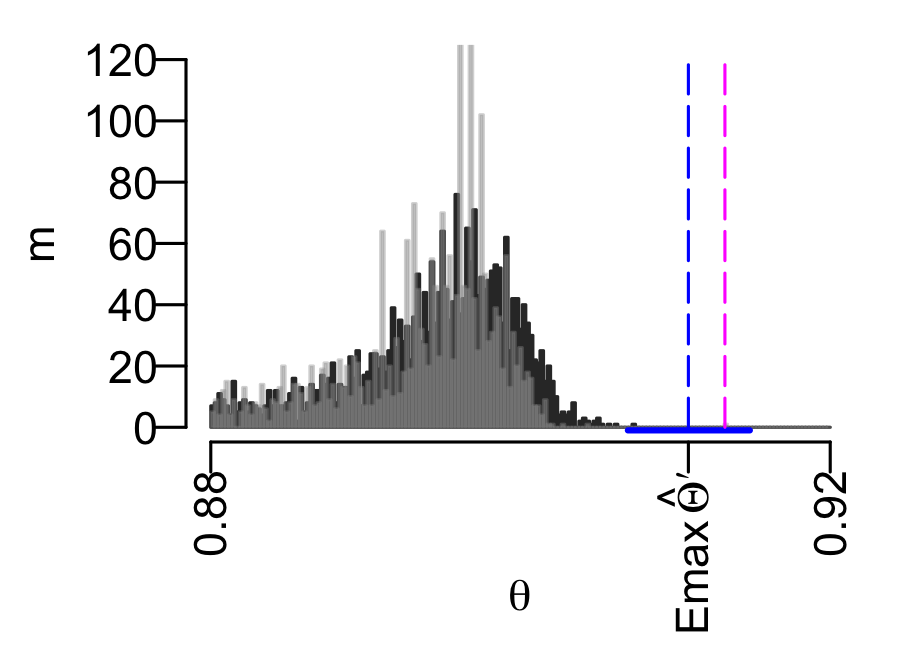}
         \caption{A realisation from a), the expected sample maximum and its CI in blue, $\hat{\theta}_{\max}$ in magenta for reference}
         \label{fig:casava_direct_bootstrap}
     \end{subfigure}
     \caption{Cassava Leaf observed performances, and a corresponding realisation, setting $\Theta' = \hat \Theta$.}
    \label{fig:casava}
\end{figure}

The three top performances appear as outliers: 0.9132, 0.9043 and 0.9028 as seen in Fig.~\ref{fig:casava_kaggle}.
Several explanations for the outlying performances have been put forward, and the interested reader can take a deep-dive in the comment section of the challenge. 
Since this is not the topic of the present work, we will not engage in the discussion here. 
The CI of $\maxtheta$ is $(0.9086,0.9177)$, no other performance is within this interval.
Already here we can decline a multiplicity-induced bias as a likely explanation. 
If we set $\Theta' = \hat{\Theta}$, we get $\text{E}\max (\hat{\Theta}') < \maxtheta$, and although the upper tail is stretched, its tip does not surpass $\maxtheta$, as illustrated in Fig.~\ref{fig:casava_direct_bootstrap}.

\subsubsection*{Melanoma Classification}
\label{sec:Melanoma}

The \textit{SIIM-ISIC Melanoma Classification} challenge \cite{Zawacki2020} consisted of more than $40,000$ images of skin lesions from various sites, split into training ($33,126$), validation ($3,295$) and test sample ($7,687$).
The data set is highly imbalanced, with less than $2\%$ melanomas, 
which reflects the reality in many cancer detection situations. 
A full description can be found in \cite{Rotemberg2021}.
The size of the test sample is smaller than the recommendations  \cite{Roelofs2019}, but reflects the reality of benchmark datasets and public challenges \cite{Willemink2020}.
We want to emphasise that the shortcomings of this challenge are not unique, and that is precisely why it serves as a useful example. 

\begin{figure}[t!]
     \centering
     {\bf Melanoma: $\hat \Theta$, $\Theta'$ and a corresponding realisation}
     \begin{subfigure}[t]{0.475\textwidth}
         \centering
        \includegraphics[width=\textwidth]{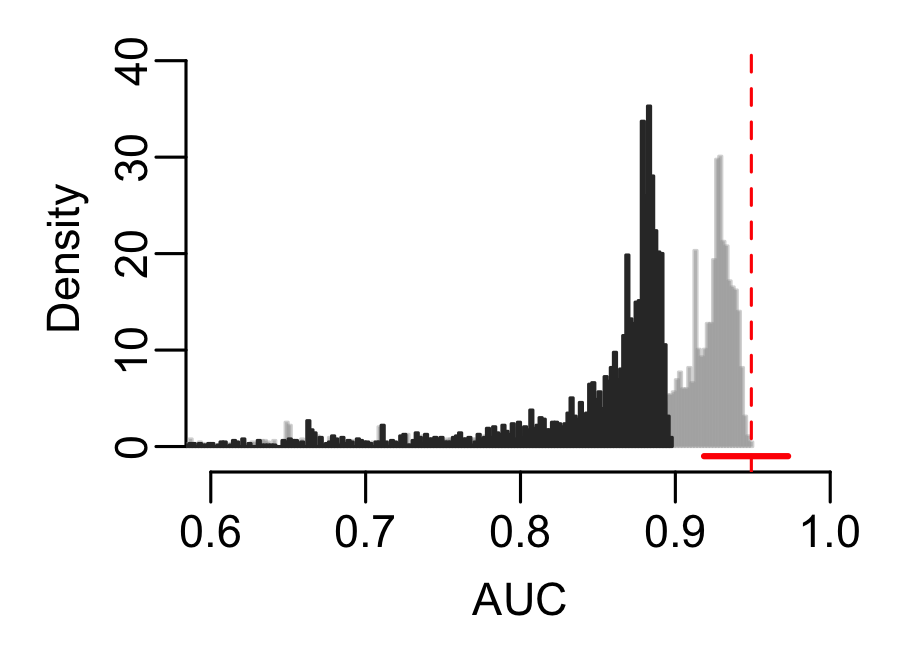}
         \caption{The JS \textit{max}-shrunk data, $\Theta'$ (black) and the observed AUC's $\hat{\Theta}$ (gray) for reference. $ \hat{\theta}_{\max}$ and its CI in red.}
         \label{fig:melanoma_bootstrap_cropped}
     \end{subfigure}
     \hfill
     \begin{subfigure}[t]{0.475\textwidth}
         \centering
         \includegraphics[width=\textwidth]{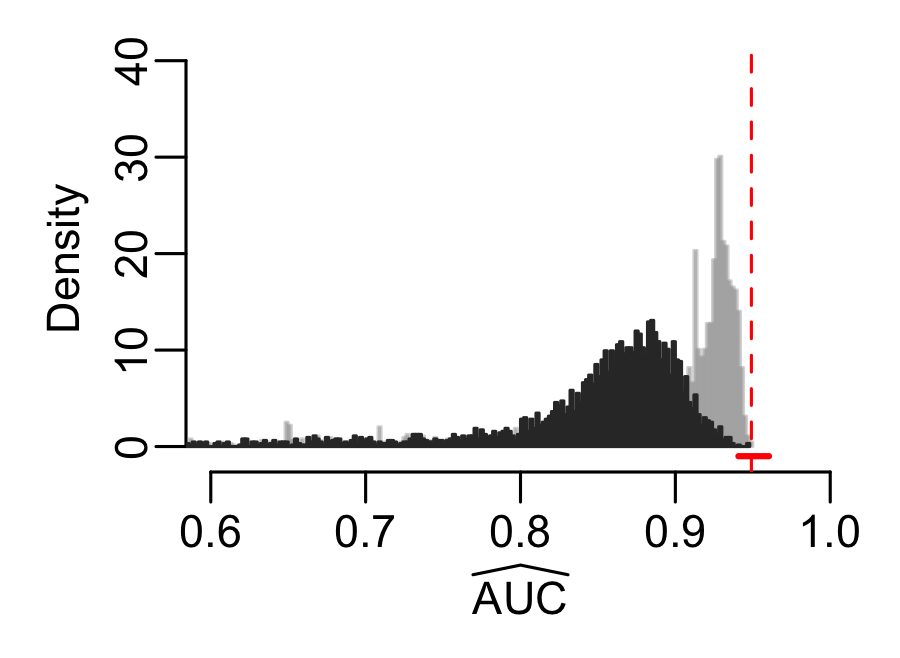}
         \caption{A realisation from a), $\text{E} \max (\hat{\Theta}') =\hat{\theta}_{\max}$ and its CI.}
         \label{fig:melanoma_realisation_cropped}
     \end{subfigure}
     \caption{The JS \textit{max}-shrunk AUCs of the Melanoma challenge. In the shrinkage distribution, the maximum AUC is $0.8991$ which leads to the expected sample maximum AUC of 0.9490. AUCs below $0.6$ not shown.}
    \label{fig:melanoma_auc_cropped}
\end{figure}

The prize money was set to $10,000$ USD for the best-performing classifier, measured as AUC on the test sample. 
Fig.~\ref{fig:melanoma_bootstrap_cropped} shows the results.
The team `All Data Are Ext', consisting of three Kaggle Grandmasters, won the challenge, achieving an AUC of $0.9490$
, and their method and strategy is explained in \cite{Ha2020}. 
Although we show that the AUC is an optimistic estimate for the melanoma detection SOTA, we do not question their rank in the challenge. 
Quite the contrary; the winning team chose many of the verified strategies from Section~\nameref{sec:RelatedWork}; use of additional data from previous years' challenges, data augmentation, ensemble learning as a strategy to avoid overfitting, implicitly supported by \cite{Talagrand1996}, training on multiple classes, shown by \cite{Feldman2019} to slow down the overfitting.

The observed maximum is an unlikely SOTA candidate, 
and to give a better estimate of the SOTA for the Melanoma challenge, we shrink according to Eq.~\ref{eq:JamesStein}.

Fig.~\ref{fig:melanoma_bootstrap_cropped} shows the JS \textit{max}-shrunk AUCs, with maximum $0.8991$. 
Fig.~\ref{fig:melanoma_realisation_cropped} shows a corresponding sample, where the maximum is close to $0.9490$.
There are 2013 teams with performance above the new SOTA candidate. 
The discrepancy between the sample maximum and the new SOTA candidate is big and has big implications. 
From an AUC close to $0.95$ to barely $0.90$ is a substantial drop in performance.
Almost two thirds of the classifiers have performances above the new SOTA estimate. 

As was shown in Table~\ref{tab:noniid}, illustrated in Fig.~\ref{fig:bias_thetamin_rho}, the bias is reduced for correlated classifiers, so the analysis assuming independence overestimates the bias.
In this challenge, we can assume multiple sources of dependency: Access to the same training set, models pre-trained on the same data, similar models, etc.
However, the wide CI for a single classifier tells the story of an estimator with large variance, and hence the bias of the sample maximum will nevertheless be substantial.

\section*{Discussion} 
\label{sec:Discussion}


We believe that multiplicity in benchmark performance has been neglected mainly because of the lack of formalization: What is meant by state-of-the-art, and what question does it answer?
Our main contribution has been to phrase this question (Eq.~\ref{def:SOTA}) and emphasize the distinction between best performance in an ensemble, and performance of best classifier. 
Ignoring this distinction, and thus the multiplicity effect, lets you invent a coin-flip classifier with remarkable results (Section~\nameref{sec:Multiple}).
While this seems like a toy example, it does not differ from public challenges.


The beauty of the approach presented in this work - that we do not even have to know \textit{which} classifier is the best to estimate the best performance - becomes the curse for the public challenge: We do not know which team is the best, only which was good combined with luck,
or \textit{uncertainty} as it would be named in statistics.


We must keep in mind 
that simpler models tend to overestimate the bias. 
It is obvious from Fig.~\ref{fig:obesity_shrunk_for_expect_realisation} that the chosen approach to find a good empirical distribution is not optimal.
A sample from the empirical distribution in Fig.~\ref{fig:obesity_shrunk_for_expect} should ideally result in a sample similar to the observed sample estimates, that is, an overlap between the black and the gray histograms in Fig.~\ref{fig:obesity_shrunk_for_expect_realisation}. 
That would require more parameters and more tweaking.
When deep-diving into a specific problem, it can be worthwhile spending some extra time on that.


There are several approaches that can contribute to the problem of multiplicity, as presented in Section~\nameref{sec:RelatedWork}.
Common for the solutions presented there are that they require additional data or more elaborate set-ups for public challenges.
The approach presented here offers a tool that can be applied to already published results. 

\subsubsection*{Dependency models}
The dependency model discussed in Section~\nameref{sec:id_dep} is simple, and does not fully reflect the complex dependencies between machine learning models - from using the same architectures, training on the training sample released for the specific challenge, and adjusting parameters to the validation sample.
A hierarchical model with more conditional layers can reflect this, but it will also require more parameters for the user to choose or estimate from the data.  
A simpler model often outperforms a more complex one, where more parameters need to be estimated - a well-known and difficult balancing exercise. 
The biggest simplification is having equal correlation coefficients.
Top-performing classifiers are likely to have higher correlation, than between high- and low-performing models, but not necessarily.
An investigation into pairwise correlation of classifiers requires the release of results for the entire data vector, not only the for the test sample as a whole.

For a specific challenge, and a specific SOTA estimation, it is well worth going into a deeper investigation, taking into account the large impact correlation has on bias, as demonstrated in Table~\ref{tab:noniid}.

\subsubsection*{JS weight}
Apart from using the generalized JS estimators, a natural way forward would be to have individual weights $w_j$ instead of $\mathbf{w} = w$.
The ultimate goal of the estimation is not $\Theta$, but only $\max\Theta$. 
A first (heuristic) approach could be to let the individual weights reflect the distance the sample maximum, as those further away are less likely to be contributing.

The JS estimator is vulnerable to outliers, as illustrated for the Cassava leaf data. 
As with any estimator, one should consider its drawbacks before engaging it.
Shrink estimators represent a risk to outlying performances, but remarkable results will draw substantial attention from the machine learning community to avoid a blind trust in results and analysis, as shown in the Cassava example.

\nolinenumbers

%
%

\bibliography{plos_bibtex_sample}

\end{document}